\documentclass[nopubldata]{lpor2014}
\graphicspath{{figs/}}
\usepackage{placeins}
\usepackage{blindtext}

\usepackage{hyperref}
\hypersetup{%
  colorlinks,
  plainpages=false,
  pdfpagelabels,
  breaklinks,
  pdfview=Fit,
  bookmarksopen,
  bookmarksnumbered,
  linkcolor=black,
  anchorcolor=black,
  citecolor=black,
  filecolor=black,
  urlcolor=LPRred
  }%

\usepackage{subfigure,amsmath,amssymb,graphicx,epstopdf,xcolor}
\usepackage{textcomp}
\usepackage{enumerate}
\usepackage{placeins} 
\usepackage[sort,numbers,sort&compress]{natbib}

\begin{document}
%

\abstract{Entangled photon pairs generated within integrated devices must often be spatially separated for their subsequent manipulation in quantum circuits. Separation that is both deterministic and universal can in principle be achieved through anti-coalescent two-photon quantum interference. However, such interference-facilitated pair separation (IFPS) has not been extensively studied in the integrated setting, where the strong polarization and wavelength dependencies of integrated couplers -- as opposed to bulk-optics beamsplitters -- can have important implications for performance beyond the identical-photon regime. This paper provides a detailed review of IFPS and examines how these dependencies impact separation fidelity and interference visibility. Focus is given to IFPS mediated by an integrated directional coupler. The analysis applies equally to both on-chip and in-fiber implementations, and can be expanded to other coupler architectures such as multimode interferometers. When coupler dispersion is present, the separation performance can depend on photon bandwidth, spectral entanglement, and the linearity of the dispersion. Under appropriate conditions, reduction in the separation fidelity due to loss of non-classical interference can be perfectly compensated for by classical wavelength demultiplexing effects. This work informs the design as well as the performance assessment of circuits for achieving universal photon pair separation for states with tunable arbitrary properties.}

\title{Deterministic separation of arbitrary photon pair states in integrated quantum circuits}
%
\author{Ryan P. Marchildon\inst{1} and Amr S. Helmy\inst{1,2}}
%
\authorrunning{R. P. Marchildon and A. S. Helmy}
%
\institute{%
   The Edward S. Rogers Department of Electrical and Computer Engineering,
   University of Toronto, 10 King\textquoteright{}s College Road, Toronto,
   Ontario M5S 3G4, Canada.
\and
   Institute for Optical Sciences, University of Toronto, 60 St. George Street, Toronto, Ontario M5S 3G4, Canada.
}
%
\mail{\email{a.helmy@utoronto.ca}}
%
%
\maketitle

\section{Introduction}
\label{Sec:Introduction}

\par By harnessing fundamental quantum properties of light such as entanglement and superposition, quantum photonic technology has enabled new paradigms in secure communications \cite{Lo_Science_1999}, quantum simulation \cite{ Ma_NaturePhysics_2011}, and enhanced metrology \cite{Giovannetti_NaturePhotonics_2011}. Mapping quantum photonic technologies into an integrated on-chip setting has become an important task for overcoming the severe stability and scalability limitations of bulk-optics implementations. Recent efforts have demonstrated on-chip quantum state generation \cite{Ghali_NatureComms_2012, Davanco_APL_2012, Horn_SciRep_2013}, manipulation \cite{Politi_Science_2008, Matthews_NaturePhotonics_2009, Sansoni_PRL_2010, Shadbolt_NaturePhotonics_2011, Wang_OptComms_2014, Silverstone_NaturePhotonics_2014} and detection \cite{Reithmaier_SciRep_2013} across numerous material platforms including GaAs \cite{Horn_SciRep_2013, Reithmaier_SciRep_2013, Wang_OptComms_2014}, silicon wire \cite{Davanco_APL_2012, Silverstone_NaturePhotonics_2014}, silica-on-silicon \cite{Politi_Science_2008, Matthews_NaturePhotonics_2009,  Shadbolt_NaturePhotonics_2011},  lithium niobate \cite{Jin_PRL_2014} and borosilicate glass \cite{Sansoni_PRL_2010}.

\par
\begin{figure}[t!]
\centering
\includegraphics[width=0.80\columnwidth]{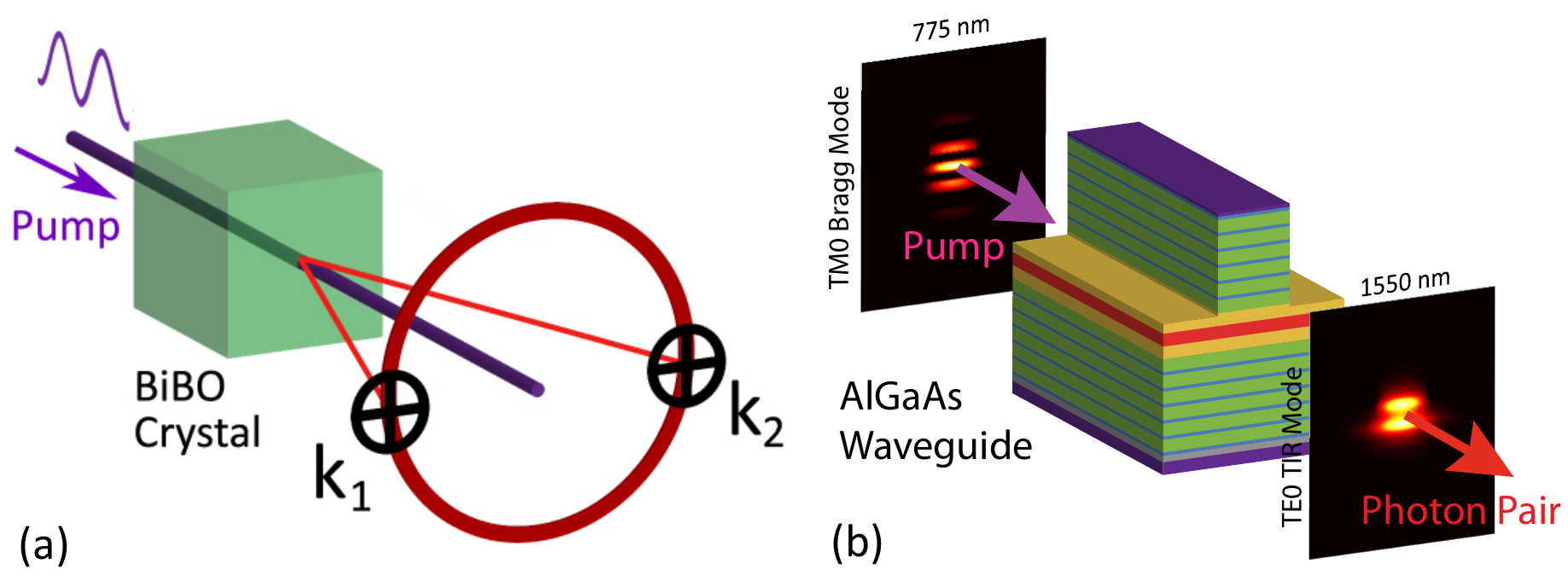}
\caption{(a) Example of photon pair generation from a bulk nonlinear crystal, where momentum conservation constrains the photons to be found only at antipodal points of the conical cross-section. Strategic collection leads to the photons deterministic separation~\cite{Matthews_NaturePhotonics_2009}. (b) Illustration of a waveguide photon pair source, based on the device architecture from Ref.~\cite{Horn_SciRep_2013}.} 
\label{Fig:IntroIllustrations}
\end{figure}

 Correlated photon pairs are an important resource for quantum photonics that can be generated on-chip by quantum dots \cite{Ghali_NatureComms_2012} or integrated nonlinear waveguides \cite{Davanco_APL_2012, Horn_SciRep_2013, Jin_PRL_2014}. As well as being both compact and efficient, integrated photon pair sources have also shown unprecedented versatility in tailoring the properties of the generated twin-photon state through dispersion engineering and birefringence management, thereby establishing control over the spectral and polarization entanglement \cite{Zhukovsky_PRA_2012, Eckstein_PRL_2011, Kang_JOSAB_2014}, photon bandwidths \cite{Abolghasem_OL_2009}, and degree of non-degeneracy. A single device can be designed to produce a variety of quantum states. These states can be selected through the pump polarization or wavelength, allowing for in-situ toggling between cross- and co-polarized generation \cite{Zhukovsky_OL_2011} in addition to continuous tunability over the photon central wavelength separation (i.e. non-degeneracy) \cite{Horn_SciRep_2013}. However, the division of generated twin-photons into different waveguides for independent manipulation can be challenging for these structures since, unlike in bulk-optics, the spatial modes available for pair production generally overlap and co-propagate with no way to separate the photons based on their spatial distribution. This contrast is illustrated in Figure~\ref{Fig:IntroIllustrations}. 

\par Ideally, pair separation should be `deterministic' so that the two photons are made to propagate in different waveguides with near-unity probability. Conventional deterministic separation methods such as wavelength-demultiplexing or polarizing beamsplitters may be unsuitable when the photons spectrally overlap, have highly tunable properties, or when entanglement in one or more degrees of freedom must be preserved and kept uncorrelated with the output path. An effective approach to pair separation is to exploit non-classicality in the two-photon statistics. One of the most familiar manifestations of such non-classicality is the Hong-Ou-Mandel (HOM) effect \cite{Hong_PRL_1987, Ou_PRL_1988}, wherein an anti-bunched state coalesces into a bunched state. In much the same way, quantum interference between two identical sources of photon pairs can be used to achieve the reverse effect: the anti-coalescence of bunched states into anti-bunched states. Such interference-facilitated pair separation (IFPS) was realized first in fiber Sagnac loops \cite{Chen_PRA_2007, Li_PRA_2009, Zhao_OL_2012} and more recently on-chip \cite{Silverstone_NaturePhotonics_2014, Jin_PRL_2014} by coherently pumping two photon pair sources, e.g. denoted $A$ and $B$, to create an approximate NOON-type state of the form $\left\vert \Psi \right\rangle = \left\vert \psi \right\rangle_{A}\left\vert 0 \right\rangle_{B} + e^{i\theta}\left\vert 0 \right\rangle_{A}\left\vert \psi \right\rangle_{B}$ where $\theta$ represents a stable relative phase, $\left\vert \psi \right\rangle$ a photon pair, and $\left\vert 0 \right\rangle$ the vacuum. Evolving this state with $e^{i\theta} = 1$ through an ideal 50:50 mode coupler leads to an anti-bunched output so long as no `which-way' information is available to distinguish $\left\vert \psi \right\rangle_{A}$ from $\left\vert \psi \right\rangle_{B}$. This holds true even if the photons comprising $\psi$ are themselves distinguishable \cite{Kim_JOSAB_2005}. Since this interference is fundamentally a Feynman-path phenomenon, IFPS with an ideal 50:50 coupler does not rely on the \textit{intrinsic} properties of the photon pair, but on only the indistinguishability of the sources \cite{Pittman_PRL_1996, Kim_JOSAB_2005}. Hence, IFPS has the potential to provide universal deterministic separation independent of the photon properties, allowing a single monolithically integrated device to separate any arbitrary two-photon state $\left\vert \psi \right\rangle$ into different waveguides.

\par A variety of integrated couplers are available for mediating on-chip quantum interference, among which the directional coupler has been the most widely-used \cite{Politi_Science_2008, Matthews_NaturePhotonics_2009, Sansoni_PRL_2010, Shadbolt_NaturePhotonics_2011, Wang_OptComms_2014}. However, unlike the bulk-optics beamsplitters, the power splitting ratio of a directional and other integrated coupler types can have significant wavelength and polarization dependence. Although these splitting ratio dependencies have evidenced little impact on recent HOM-type experiments, where the photons were co-polarized and approximately degenerate, their implications cannot be neglected for the more general states to which IFPS may be applied, which include states with non-identical photons. As more quantum circuits begin to incorporate on-chip photon pair sources, and the properties of these photons become increasingly tunable, there is a growing importance to understand IFPS and its nuances.

\section{Directional Couplers with Dispersion}
\label{Sec:Background}

\par Directional couplers have thus far been the most ubiquitous coupler type for implementing on-chip quantum interference \cite{Politi_Science_2008, Matthews_NaturePhotonics_2009, Sansoni_PRL_2010, Shadbolt_NaturePhotonics_2011, Wang_OptComms_2014}. When IFPS is implemented using such devices, its analysis has several key distinctions from the typical bulk-optics treatment. The usual assumption of a wavelength-independent splitting ratio for bulk-optics beamsplitters is not generally appropriate for integrated couplers, particularly when non-degenerate states are considered. A coupler's response may differ significantly not only between the two photons, but also across the spectral bandwidth of each photon. 

\begin{figure}[!t]
\centering
\includegraphics[width=0.80\columnwidth]{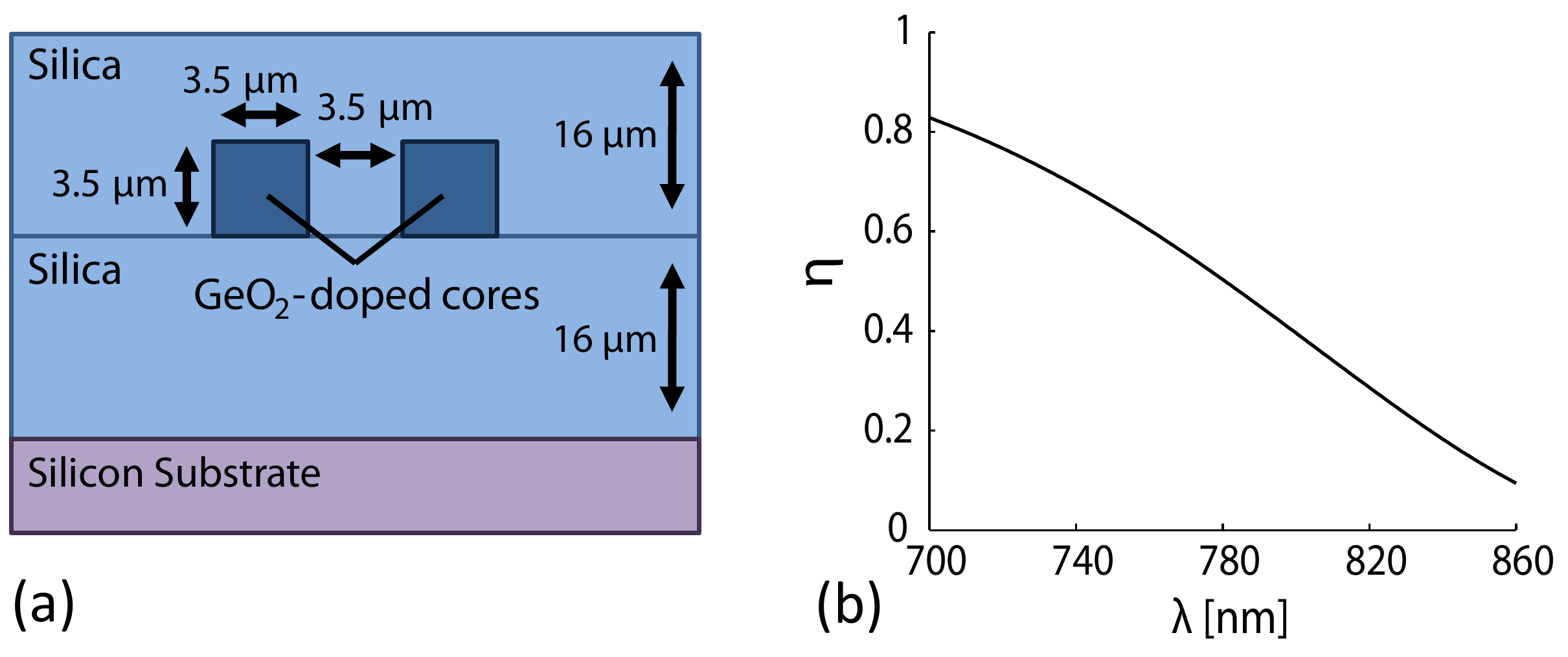}
\caption{(a) Illustrated cross-section of the simulated silica-on-silicon directional coupler based on Ref.~\cite{Thompson_IETCDS_2011}; and (b) its calculated TE splitting ratio.} 
\label{Fig_CouplerExample}
\end{figure}

\par To place this in perspective, consider the silica-on-silicon directional coupler depicted in Figure~\ref{Fig_CouplerExample}(a). This coupler design was used in several early examples of on-chip quantum interference exhibiting high HOM visibilities (Refs.~\cite{Thompson_IETCDS_2011, Politi_Science_2008, Matthews_NaturePhotonics_2009, Shadbolt_NaturePhotonics_2011}). Figure~\ref{Fig_CouplerExample}(b) shows its numerically simulated splitting ratio $\eta(\lambda)$ for a range of wavelengths around $\lambda = \textrm{780~nm}$.  The variation in $\eta$ is significant. Many waveguide-based twin-photon sources can have their photon wavelengths non-degenerately tuned over such a range \cite{Horn_SciRep_2013}, or can generate large-bandwidth states \cite{Abolghasem_OL_2009} of comparable span, and hence these changes in $\eta$ cannot be neglected. Their impact on IFPS is analysed in Section~\ref{Sec:IFPS_ImpactOfCouplerAttributes}.

\par In the absence of modal mismatch, the evanescent coupling of two waveguides leads to a general splitting ratio dependence of $\eta_{\sigma}(\omega) = \cos^{2}\left(L \kappa_{\sigma}(\omega) \right)$ in frequency space \cite{Yariv_JQE_1973}, where $\kappa_{\sigma}(\omega)$ represents the coupling strength, $L$ is the coupling interaction length, and $\sigma \in \{\mathrm{TE},\mathrm{TM} \}$ is the polarization. As defined here, $\eta_{\sigma}(\omega) = 1$ corresponds to conditions of zero waveguide power transfer.

\subsubsection*{Mode Operator Transformations}

\par All couplers discussed in this manuscript are assumed to be symmetric such that modal mismatch is negligible. If $\eta$ is taken as dispersionless, then quantum mode operators are transformed using
\begin{equation}
\renewcommand{\arraystretch}{1.25} \left[ \begin{array}{c} \hat{b}_{\sigma}^{A \dagger}(\omega) \\ \hat{b}_{\sigma}^{B \dagger}(\omega) \end{array} \right] = \renewcommand{\arraystretch}{1.25} \left[ \begin{array}{cc} \sqrt{\eta} & i\sqrt{1-\eta} \\  i\sqrt{1-\eta}  & \sqrt{\eta} \end{array} \right] \renewcommand{\arraystretch}{1.25}  \left[ \begin{array}{c} \hat{a}_{\sigma}^{A \dagger}(\omega)  \\ \hat{a}_{\sigma}^{B \dagger}(\omega) \end{array} \right],
\label{Eqn:SimpleTransformation}
\end{equation}
where $\hat{b}^{A \dagger}_{\sigma}(\omega)$ and $\hat{b}^{B \dagger}_{\sigma}(\omega)$ denote the mode operators at the coupler output, which differ from those at the input ($\hat{a}^{A \dagger}_{\sigma}(\omega)$ and $\hat{a}^{B \dagger}_{\sigma}(\omega)$) by a phase shift. However, when fully incorporating coupler dispersion, this transformation becomes
\begin{equation}
\renewcommand{\arraystretch}{1.25} \left[ \begin{array}{c} \hat{b}^{A \dagger}_{\sigma}(\omega) \\ \hat{b}^{B \dagger}_{\sigma}(\omega) \end{array} \right] = \renewcommand{\arraystretch}{1.25} \left[ \begin{array}{cc} \cos\left(\kappa_{\sigma}(\omega)L\right) & i\sin\left(\kappa_{\sigma}(\omega)L\right) \\  i\sin\left(\kappa_{\sigma}(\omega)L\right)  & \cos\left(\kappa_{\sigma}(\omega)L\right) \end{array} \right] \renewcommand{\arraystretch}{1.25}  \left[ \begin{array}{c} \hat{a}^{A \dagger}_{\sigma}(\omega)  \\ \hat{a}^{B \dagger}_{\sigma}(\omega) \end{array} \right].
\label{Eqn_ModeTransformation}
\end{equation}
It has been written explicitly in terms of the sinusoidal dependence on $\kappa_{\sigma}(\omega)$ to emphasize that, unlike in typical bulk-optics or non-dispersive treatments, the matrix elements can now change relative sign. For example, for $\pi/2 < \kappa_{\sigma}(\omega)L < \pi$, the diagonal elements become negative relative to the off-diagonal elements. It no longer suffices to write the transformations only in terms of the power splitting ratio $\eta(\omega)$ because these details would be lost. The effect this can have on quantum interference is similar to a $\pi$ phase shift between the interfering paths, and only occurs when coupler dispersion is present, as will be shown in Section~\ref{SubSubSec:SalientFeatures}.

\section{Interference-Facilitated Photon Pair Separation}
\label{Sec:IFPS_General}

\subsection{Existing Demonstrations}

\par The first demonstration of time-reversed HOM interference for the purpose of deterministic pair separation was reported using a nonlinear fiber Sagnac loop in 2007 by Chen, Lee, and Kumar \cite{Chen_PRA_2007}. Each propagation direction then became a source of identical photon pairs leading to quantum interference between the CW and CCW directions. Several variations of this experiment followed, focusing now on distinguishable photons (but still indistinguishable sources) \cite{Li_PRA_2009, Zhao_OL_2012}. The first on-chip implementation of IFPS involved the interference of co-polarized photon pairs generated near degeneracy from two silicon waveguides using a multi-mode interferometer \cite{Silverstone_NaturePhotonics_2014}. IFPS was later demonstrated in a periodically-poled lithium niobate (PPLN) waveguide circuit \cite{Jin_PRL_2014} using directional couplers, but only with identical photons.

\subsection{General Expressions}
\label{SubSec:General_IFPS_Expressions}

\par Studying the implications of coupler dispersion for generalized IFPS requires a formalism that can be applied to any arbitrary co-propagating photon pair state generated by an integrated source. Such a formalism is presented here in some detail. Its development follows a similar approach as the original HOM theory \cite{Hong_PRL_1987, Ou_PRL_1988}. However, several important distinctions between IFPS and the HOM effect will arise that are pertinent to the generic scope of this work and will be addressed in Section~\ref{SubSec:IFPS_HOM_differences}.

\par
\begin{figure} [t!]
\centering
\includegraphics[width=0.8\columnwidth]{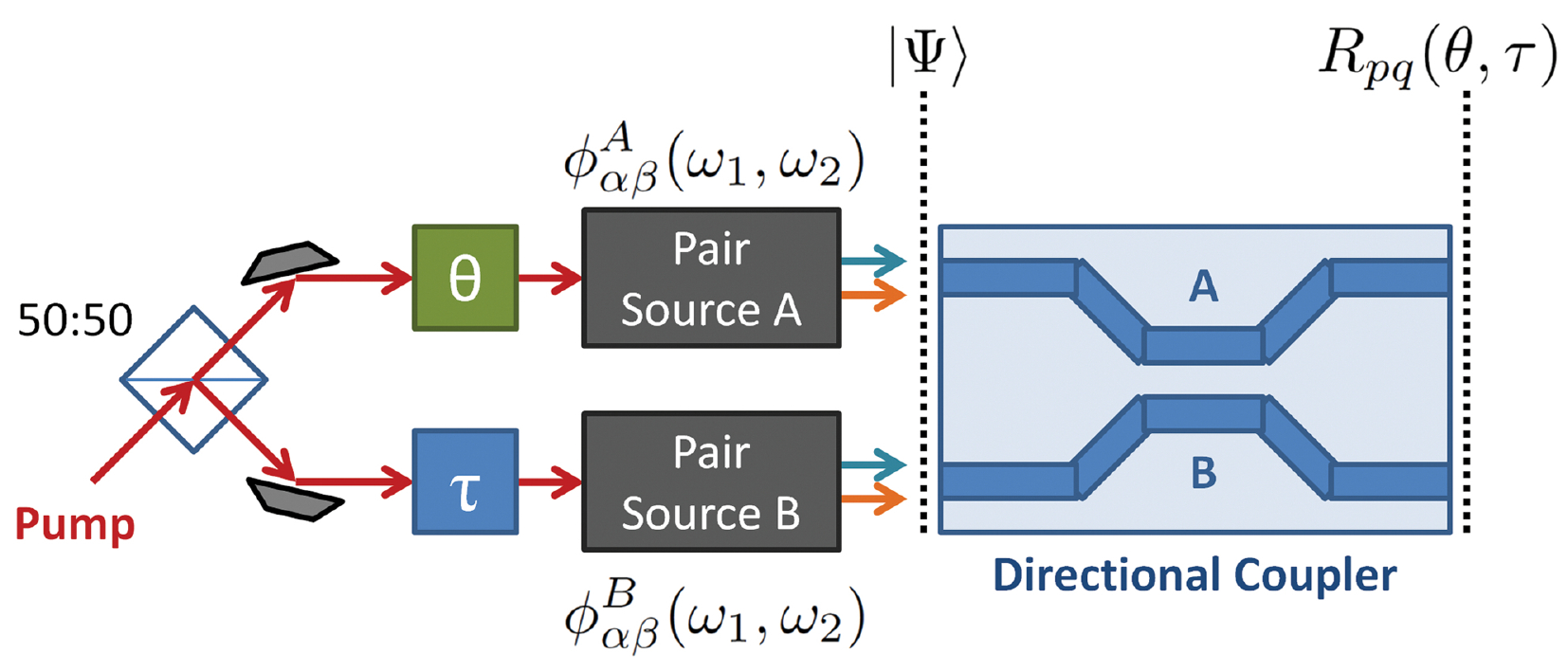}
\caption{Block-schematic of two coherently pumped photon pair sources producing the state $\left\vert \Psi \right\rangle$ of Equation~(\ref{Eqn_StartingState_FullVers}), with subsequent interference through a directional coupler.} 
\label{Fig_DetSepCircuit}
\end{figure}

 Consider two coherently pumped waveguide photon pair sources denoted $A$ and $B$ as depicted in Figure~\ref{Fig_DetSepCircuit}, and for simplicity assume all waveguides are single-mode over the bandwidths of interest. The outputs of sources $A$ and $B$ are associated with the mode creation operators $\hat{a}^{A \dagger}_{\sigma}(\omega)$ and $\hat{a}^{B \dagger}_{\sigma}(\omega)$ respectively, which satisfy the commutation relation $\big[ \hat{a}^{m}_{\alpha}(\omega), \hat{a}^{n \dagger}_{\beta}(\omega') \big] = \delta_{\alpha \beta}\delta_{m n} \delta (\omega - \omega')$. The subscript $\sigma$ refers to polarization. In the weak pumping regime, where higher-order pair production is negligible, a general expression for the generated quantum state is
\begin{align}
\left\vert \Psi \right\rangle = \sum_{\alpha \beta}& \int \mathrm{d}\omega_{1} \mathrm{d} \omega_{2} \Big[ e^{i \theta}\phi^{A}_{\alpha \beta}(\omega_{1},\omega_{2}) \hat{a}^{A \dagger}_{\alpha}(\omega_{1}) \hat{a}^{A \dagger}_{\beta}(\omega_{2}) \nonumber \\
&+  e^{-i\left[ \omega_{1} + \omega_{2}  \right] \tau}\phi^{B}_{\alpha \beta}(\omega_{1},\omega_{2}) \hat{a}^{B \dagger}_{\alpha}(\omega_{1}) \hat{a}^{B \dagger}_{\beta}(\omega_{2}) \Big] \left\vert \text{vac} \right\rangle,
\label{Eqn_StartingState_FullVers}
\end{align}
where $\alpha$ and $\beta$ denote polarization, $\vert \text{vac} \rangle$ refers to vacuum, and it is assumed that both photon pair sources are pumped with equal strength. The function $\phi(\omega_{1}, \omega_{2})$, called the joint spectral amplitude (JSA), completely describes the spectral properties of the two-photon state, including spectral entanglement. The parameter $\theta$ represents a stable relative phase, amenable to electro-optic or thermal tuning, which acquires a quiescent value of $\theta = \pi$ for sources pumped through a beamsplitter or on-chip directional coupler, and $\theta = 0$ for an on-chip Y-coupler. For generality, a time delay $\tau$ between the pairs generated by one source relative to the other has been included. The function $\phi_{\alpha \beta}^{j}(\omega_{1},\omega_{2})$ is the joint spectral amplitude (JSA) of the photon pairs generated by source $j \in \{A,B\}$. It is normalized such that $ \left\langle \Psi \vert \Psi \right\rangle = \sum_{j} \sum_{\alpha \beta} \int \mathrm{d}\omega_{1} \mathrm{d} \omega_{2} \big\vert \phi_{\alpha \beta}^{j}(\omega_{1},\omega_{2}) \big\vert^{2} = 1$ and defined relative to the same spatial coordinate as $\hat{a}_{\sigma}^{j \dagger}(\omega)$. For pair generation through SPDC from $\chi^{(2)}$ nonlinearities, or spontaneous four-wave mixing (SFWM) from $\chi^{(3)}$ nonlinearities, $\phi_{\alpha \beta}^{j}(\omega_{1},\omega_{2})$ is determined by the parameters of the nonlinear interaction \cite{Chen_PRA_2005, Yang_PRA_2008}.

\begin{figure*}[!htb]
     \parbox[b]{0.55\textwidth}{
         \includegraphics[width=0.53\textwidth]{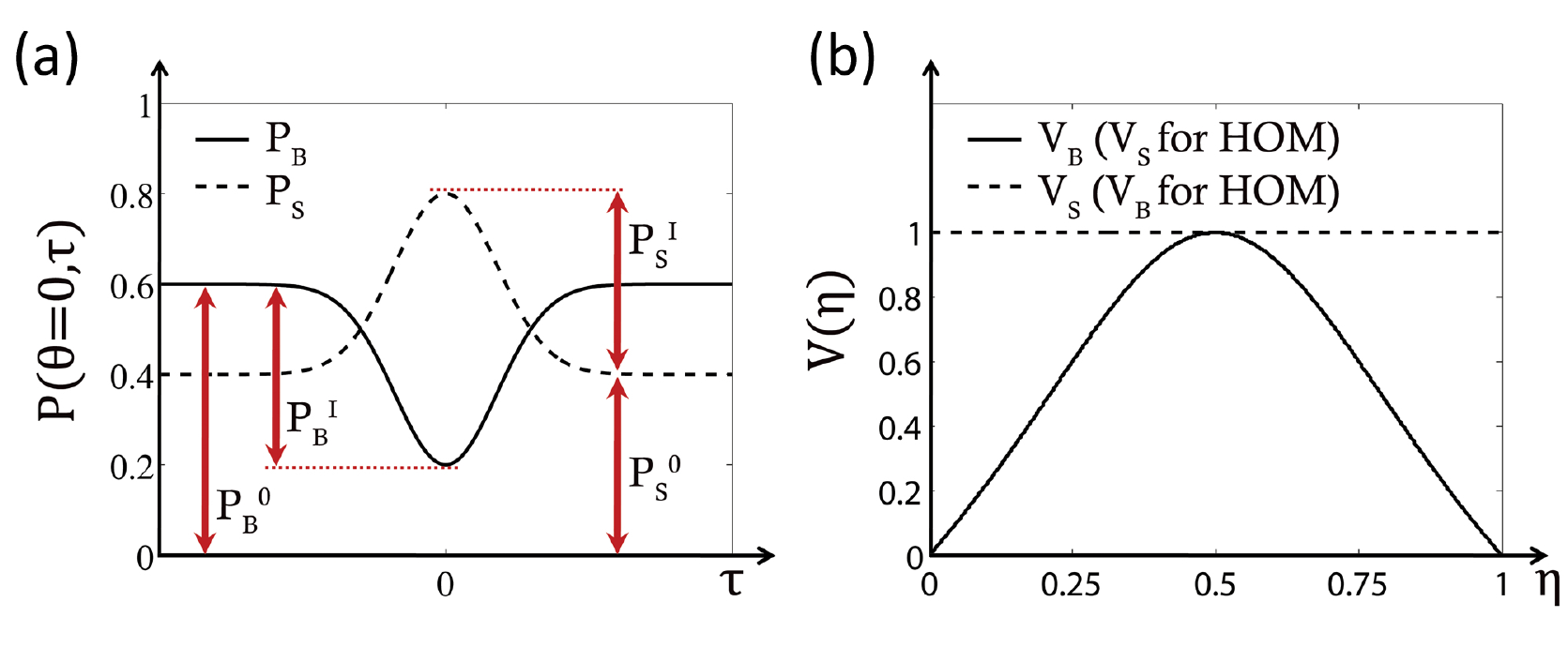}
     }\hfill
     \parbox[b]{0.45\textwidth}{
         \caption{(a) Bunched and anti-bunched (separated) outcome probabilities as related to their `classical' and `interference' components; shown at $\theta=0$ for perfect path indistinguishability and a constant splitting ratio of $\eta=0.276$. (b) Comparison of IFPS interference visibility $\eta$-dependence when the splitting ratio is a constant; the relation to HOM visibilities is also indicated.} 
         \label{Fig:Schematic2}
     }
\end{figure*}

\par The state $\vert \Psi \rangle$ evolves through a directional coupler with splitting ratio $\eta_{\sigma}(\omega)$ and transforms according to Equation~(\ref{Eqn_ModeTransformation}). It is assumed to remain in a pure state throughout this evolution. A deterministically-separated state, by definition, would always lead to a coincidence detection for a pair of detectors placed at the coupler outputs, if photon losses, detector efficiencies and dead-times are neglected. While these non-idealities are inevitable in practice, performance may nonetheless be quantified through a theoretical (ideal) probability for finding one photon in waveguide $p$ and the other in waveguide $q$. This probability can be expressed as 
\begin{equation}
R_{pq}(\theta, \tau) = R^{0}_{pq} + \cos(\pi \delta_{pq}) R^{\text{I}}_{pq}(\theta, \tau),
\label{Eqn_CountRate}
\end{equation}
where $\delta_{pq}$ is the Kronecker delta, 
\begin{align}
R^{0}_{pq}  = \sum_{\alpha \beta}  \int  \mathrm{d} \omega_{1} \mathrm{d} \omega_{2} \, \Big( &\big\vert \Phi^{A \rightarrow pq}_{\alpha \beta}(\omega_{1}, \omega_{2}) \big\vert^{2}   \nonumber \\
+  &\big\vert \Phi^{B \rightarrow pq}_{\alpha \beta}(\omega_{1}, \omega_{2}) \big\vert^{2} \Big),
\label{Eqn_R0}
\end{align}
represents the classical probability contributions from sources $A$ and $B$ in the absence of interference, and
\begin{align}
R^{\text{I}}_{pq}(\theta, \tau) =  \sum_{\alpha \beta} \int & \mathrm{d} \omega_{1} \mathrm{d} \omega_{2} \,   2\text{Re} \Big\{ e^{-i\theta} \Phi^{B \rightarrow pq}_{\alpha \beta}(\omega_{1}, \omega_{2}) \nonumber \\
&\times \Phi^{*A \rightarrow pq}_{\alpha  \beta}(\omega_{1},\omega_{2}) e^{-i(\omega_{1} + \omega_{2})\tau} \Big\},
\label{Eqn_RI}
\end{align}
describes the non-classical influence of path interference, where the superscripts $j \rightarrow pq$ are used in making the definitions
\begin{equation}
\Phi^{j \rightarrow pq}_{\alpha \beta}(\omega_{1}, \omega_{2}) = \phi^{j}_{\alpha \beta}(\omega_{1}, \omega_{2}) G^{j \rightarrow p}_{\alpha}(\omega_{1}) G^{j \rightarrow q}_{\beta}(\omega_{2}),
\label{Eqn_Phi}
\end{equation}
and
\begin{equation}
G^{j \rightarrow q}_{\sigma}(\omega) = \begin{cases} \cos{\left(\kappa_{\sigma}(\omega)L\right)}, & \mbox{if } j = q \\ \sin{\left(\kappa_{\sigma}(\omega)L\right)}, & \mbox{if } j \neq q \end{cases}.
\label{Eqn_F}
\end{equation}
These expressions are normalized such that
\begin{equation}
\sum_{pq}R_{pq}(\theta,\tau) = 1.
\end{equation}
The total probability for obtaining an anti-bunched (i.e. separated) outcome is 
\begin{equation}
P_{\mathrm{S}}(\theta,\tau)  \equiv P_{\mathrm{S}}^{0} + P_{\mathrm{S}}^{I}(\theta, \tau) = R_{AB}(\theta,\tau) + R_{BA}(\theta,\tau), 
\end{equation}
with classical and non-classical contributions given by $P_{\mathrm{S}}^{0} = R_{AB}^{0} + R_{BA}^{0}$ and $P_{\mathrm{S}}^{I}(\theta, \tau) = R_{AB}^{I}(\theta, \tau) + R_{BA}^{I}(\theta, \tau) $ respectively. Similarly, the total probability of obtaining a bunched (i.e. non-separated) outcome is given by $P_{\mathrm{B}}(\theta,\tau) = R_{AA}(\theta,\tau) + R_{BB}(\theta,\tau) = 1 - P_{\mathrm{S}}(\theta,\tau) $, which can likewise be defined in terms of $P_{\mathrm{B}}^{0}$ and $P_{\mathrm{B}}^{I}(\theta, \tau)$ contributions. These definitions have been illustrated in Figure~\ref{Fig:Schematic2}(a). Note that the classical contributions obey the constraint $P_{\mathrm{S}}^{0} + P_{\mathrm{B}}^{0} = 1$, while the non-classical contributions obey the equality $\left\vert P_{\mathrm{S}}^{I} \right\vert  = \left\vert P_{\mathrm{B}}^{I} \right\vert$ so constructive interference in the anti-bunched amplitudes will be balanced by destructive interference in the bunched amplitudes and vice-versa.

\par The interference contribution $P_{\mathrm{S}}^{I}(\theta, \tau)$ assumes its maximal amplitude when: 
\begin{enumerate}[(i) ]

\item  $\theta$ is a multiple of $\pi$; \\

\item the sources are indistinguishable so that \\ $\Phi_{\alpha \beta}^{A \rightarrow pq}(\omega_{1}, \omega_{2}) = \Phi_{\alpha \beta}^{B \rightarrow pq}(\omega_{1}, \omega_{2})$;\\

\item  and $\exp(-i[\omega_{1} + \omega_{2}]\tau)$ is a constant so that all frequency components interfere in-phase, which is satisfied when $\tau = 0$. 

\end{enumerate}
As $\theta$ is varied, the photon behaviour oscillates between perfect bunching $(P_{\mathrm{B}} \rightarrow 1)$ and perfect anti-bunching $(P_{\mathrm{S}} \rightarrow 1)$ every $\pi$ radians. The ratio of non-classical to classical contributions towards the outcome probabilities is quantified by the ideal interference visibilities
\begin{equation}
V_{\mathrm{S}} = \frac{\left\vert P_{\mathrm{S}}^{I}(\theta, \tau = 0) \right\vert}{P_{\mathrm{S}}^{0}},  \,\,\,\,\,\,\,\,\,\,\,\,\,\,\,\, V_{\mathrm{B}} = \frac{\left\vert P_{\mathrm{B}}^{I}(\theta, \tau = 0) \right\vert}{P_{\mathrm{B}}^{0}},
\label{Eqn_InterferenceVisibility}
\end{equation}
which are bounded by [0,1].

\subsection{Qualitative Differences Between IFPS and HOM Interference}
\label{SubSec:IFPS_HOM_differences}

\par It is instructive to now highlight distinctions from HOM interference, unrelated to coupler dispersion, that impact the implementation and characterization of IFPS. Firstly, when a constant splitting ratio $\eta$ is considered, the $\eta$-dependence of visibilities $V_{\mathrm{S}}$ and $V_{\mathrm{B}}$ are reversed for IFPS compared to their HOM counterparts, as indicated in Figure~\ref{Fig:Schematic2}(b). It is straightforward to show from Equations~(\ref{Eqn_R0})-(\ref{Eqn_InterferenceVisibility}) that the IFPS bunched visibility for a constant $\eta$ is 
\begin{equation}
V_{\mathrm{B}} = 2\eta(1-\eta) /\left[ \eta^2 + (1 -\eta)^2\right],
\label{Eqn:IFPS_VB}
\end{equation}
whereas the anti-bunched visibility $V_{\mathrm{S}}$ is independent of $\eta$ because the $P_{\mathrm{S}}^{0}$ and $P_{\mathrm{S}}^{I}$ probability contributions scale identically. In HOM interference on the other hand, it is the anti-bunched visibility $V_{\mathrm{S}}$ that scales according to Equation~(\ref{Eqn:IFPS_VB}), while $V_{\mathrm{B}}$ remains constant.

\par Secondly, perfect HOM interference requires the JSA involved to be symmetric in its frequency arguments. The form of its interference term is \cite{Hong_PRL_1987,Ou_PRA_2006}
\begin{equation}
P_{(\textrm{HOM})}^{I} \propto \text{Re} \left\{ \int \mathrm{d} \omega_{1} \mathrm{d} \omega_{2} \, \phi(\omega_{1}, \omega_{2}) \phi^{*}(\omega_{2}, \omega_{1})   e^{-i[\omega_{2} - \omega_{1}]\tau}  \right\}  ,
\label{Eqn:PI_HOM}
\end{equation}
where interference is maximal only when the permutation symmetry $\phi(\omega_{1}, \omega_{2}) \vert \omega_{1}\rangle_{A}\ \vert \omega_{2}\rangle_{B} = \phi(\omega_{2}, \omega_{1}) \vert \omega_{1}\rangle_{A}\ \vert \omega_{2}\rangle_{B}$ is present \cite{Ou_PRA_2006}. This is generally not satisfied by cross-polarized pair generation via SPDC or SFWM, except in the special case of maximum polarization entanglement \cite{Grice_PRA_1997, Humble_PRA_2008}. For IFPS on the other hand, as seen from Equation~(\ref{Eqn_RI}), permutations of the photon frequencies $\omega_{1}$ and $\omega_{2}$ do not affect the overall path symmetry. Hence, perfect IFPS places no fundamental restrictions on the JSA itself and can be implemented on any two-photon state so long as the JSAs from each interfering path are identical. By extension, the temporal distribution given by the JSA's Fourier transform is also without restriction. This implies perfect separation can occur even in the presence of \textit{intra}-pair photon walk-off, i.e. group velocity differences between photons from the same source do not require compensation. Hence, system birefringence does not fundamentally limit IFPS aside from its impact on coupler performance.

\par
\begin{figure}[t!]
\centering
\includegraphics[width=0.8\columnwidth]{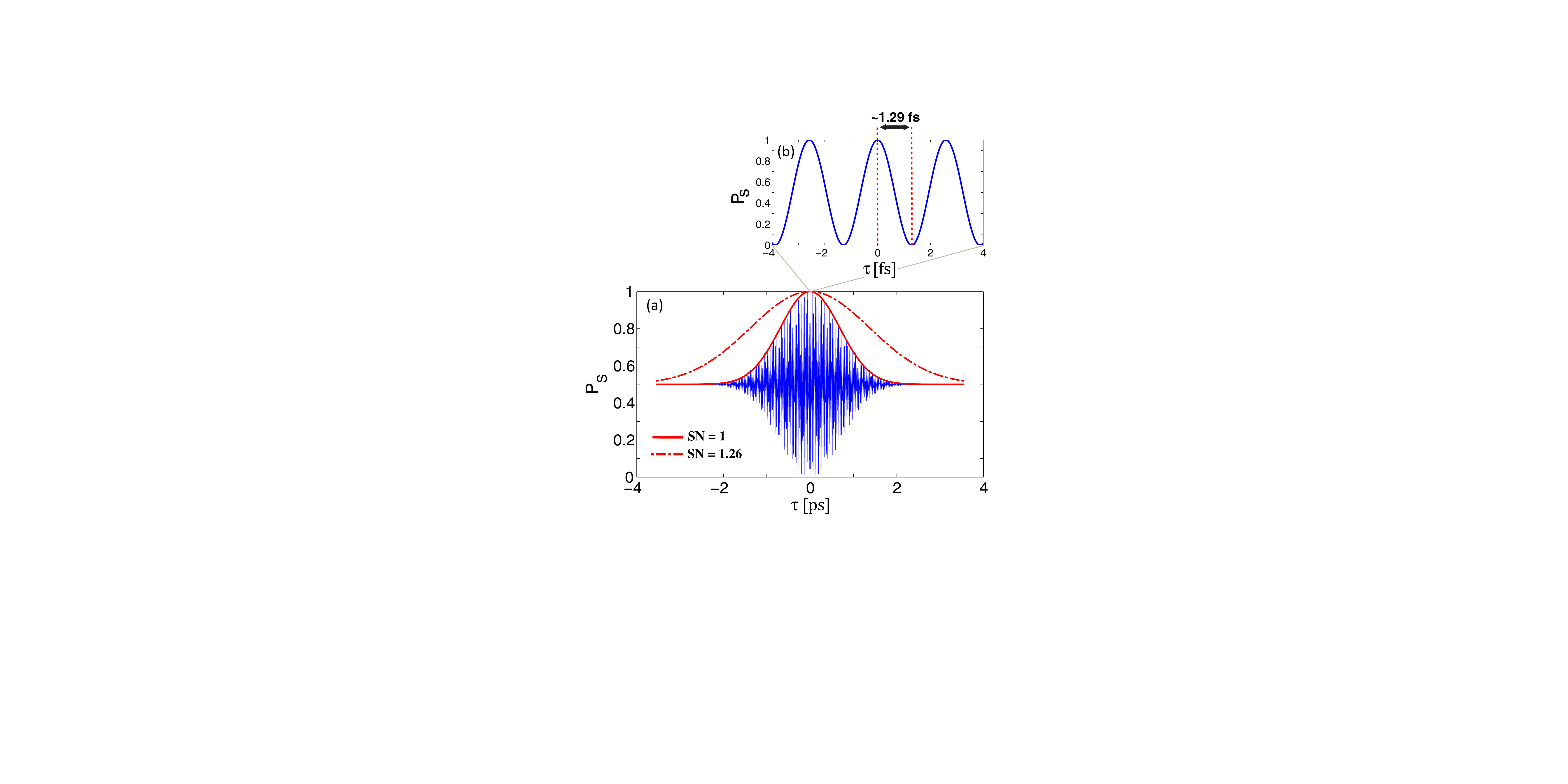}
\caption{(a) Calculated IFPS $\tau$-dependence for SPDC-generated photons at 1550~nm. The solid curve indicates the interference envelope for spectrally uncorrelated photons ($\textrm{SN} = 1$). The dashed curve shows how this envelope widens when spectral entanglement ($\textrm{SN} = 1.26$) is present. (b) Oscillatory behaviour shown at enhanced resolution to avoid aliasing.} 
\label{Fig:IFPS_TauDependence}
\end{figure}

 Thirdly, the appearance of the frequency sum $\omega_{2} + \omega_{1}$ in Equation~(\ref{Eqn_RI}), rather than the frequency difference $\omega_{2} - \omega_{1}$ obtained for the HOM treatment [Eqn.~(\ref{Eqn:PI_HOM})], can pose a challenge for stability. Recall that in HOM interference the impact of a non-zero time delay $\tau$ between interfering paths is primarily determined by the photon coherence time $\tau_{c} \propto 1/\Delta\omega$, which modulates the probability $P_{\mathrm{S}}(\tau,\theta)$ with a slowly-varying envelope \cite{Hong_PRL_1987}. Additional oscillations are observed only in the case of non-degenerate photon (or filter) central frequencies \cite{Ou_PRL_1988}. For IFPS on the other hand, the frequency sum leads to rapidly-varying oscillations with femtosecond-scale cycles, and hence a giant sensitivity to \textit{inter}-pair walk-off. As an example, consider two SPDC sources producing photons degenerate at 1550 nm. Calculations depicted in Figure~\ref{Fig:IFPS_TauDependence} show that an optical delay in one of the pump arms of merely 1.29 fs (equivalent to a free-space path difference of $\sim$ 387.5 nm) is manifested as a $\pi$ phase shift, switching the output state from anti-bunched to bunched. This stability issue is alleviated by monolithic integration, but is a serious consideration in `hybrid' cases where the coherent splitting of the pump occurs off-chip or where the outputs of two different waveguide chips are interfered through a separate directional coupler.

\par Lastly, HOM interference visibilities are commonly obtained by tracing over the entire interference envelope and comparing conditions of maximal interference ($\tau = 0$) to the classical count rate ($\vert \tau \vert > \vert \tau_{c} \vert$); however, the practicality of obtaining such a trace for IFPS is limited not only by the aforementioned oscillations, which demand extreme precision in $\tau$, but also by differences in the two-photon coherence times manifested by $\exp(-i[\omega_{2} + \omega_{1}]\tau)$. HOM coherence times are generally associated with the photon bandwidths and do not depend on frequency entanglement, whereas for IFPS the degree of spectral correlations can have a pronounced effect. In order to better appreciate this point, consider SPDC in the limiting case of a continuous-wave monochromatic pump, where the photon frequencies are perfectly anti-correlated. Since in this case the sum $\omega_{1} + \omega_{2}$ can be replaced by the monochromatic pump frequency $\omega_{p}$, a non-zero time delay $\tau$ contributes a phase factor that is now independent of $\omega_{1}$ and $\omega_{2}$, and does not degrade the coherence of the path superposition in $\vert \Psi \rangle$, hence the interference visibility never decays and the temporal width of the interference envelope approaches infinity. Realistic pump bandwidths remove the perfect anti-correlation and therefore limit the envelope width; however only in the limit of perfectly uncorrelated photons does this width approach its HOM equivalent. Figure~\ref{Fig:IFPS_TauDependence} illustrates an example of this entanglement-dependence. Spectral entanglement has been quantified using the Schmidt number ($\mathrm{SN}$) \cite{Humble_PRA_2008}, which equals one for uncorrelated states and increases with greater entanglement.

\section{Impact of Coupler Attributes on Pair Separation}
\label{Sec:IFPS_ImpactOfCouplerAttributes}

\par The implications of directional coupler dispersion for IFPS are now theoretically investigated. It is assumed that all coupler-unrelated requirements for perfect deterministic separation are met (i.e. $\tau=0$, $\phi^{A}_{\alpha \beta}(\omega_{1}, \omega_{2})= \phi^{B}_{\alpha \beta}(\omega_{1}, \omega_{2})$, and $\theta=0$ or $\pi$).

\subsection{Behaviour Near Degeneracy}
\label{SubSec:NearDegeneracy}

\par While virtually all on-chip interference experiments \cite{Politi_Science_2008, Matthews_NaturePhotonics_2009, Sansoni_PRL_2010, Shadbolt_NaturePhotonics_2011, Wang_OptComms_2014, Jin_PRL_2014} have been conducted near degeneracy, they have sampled only a small subset of the total parameter space (e.g. central wavelength separations, polarizations, photon bandwidths, coupler characteristics). The near-degeneracy regime will assume by definition that the following two approximations hold: (i) the coupling strength $\kappa_{\sigma}(\lambda)$ is locally described by a linear function in $\lambda$; (ii) the photon central wavelengths, denoted $\lambda_{1,0}$ and $\lambda_{2,0}$, are approximately equidistant from the photon degeneracy wavelength $\lambda_{\mathrm{deg}}$. The full implications of these assumptions will be elucidated in Section~\ref{SubSec:DesigningForLargeNonDeg}.

\par In order to provide a comprehensive overview of IFPS behaviour over a broad range of conditions, simple dimensionless parameters may be introduced for both the coupler response and twin-photon properties. This is achieved by replacing $L\kappa_{\sigma}(\lambda)$ in $\eta_{\sigma}(\omega) = \cos^{2}\left(L \kappa_{\sigma}(\omega) \right)$ with the dimensionless variable $\xi_{\sigma}(\lambda ; \lambda_{\mathrm{deg}})$ defined as 
 \begin{equation}
 \xi_{\sigma}(\lambda ; \lambda_{\mathrm{deg}}) = \xi^{\text{0}} +  \Delta\xi_{\sigma} + [\lambda/\lambda_{\mathrm{deg}} - 1]\mathrm{M}_{\sigma} ,
 \label{Eqn_CouplerParameterization}
 \end{equation}  
where $\xi^{\text{0}}= \pi/4 + m\pi$ ($m$ an integer) is the ideal value for a perfect 50:50 split at the degeneracy wavelength $\lambda_{\mathrm{deg}}$,
\begin{equation}
\Delta \xi_{\sigma} =  \left[ L\kappa_{\sigma}(\lambda_{\textrm{deg}}) - \pi/4, \mod \pi \right]
\end{equation}
is a systematic offset defining the true splitting ratio at $\lambda_{\mathrm{deg}}$, and 
\begin{equation}
\mathrm{M}_{\sigma} = \lambda_{\mathrm{deg}}L\left[\mathrm{d}\kappa_{\sigma}(\lambda)/\mathrm{d}\lambda \right]
\end{equation} 
characterizes the first-order coupler dispersion evaluated at $\lambda_{\mathrm{deg}}$. Since modal mismatch has been neglected, the splitting ratio $\eta_{\sigma}(\lambda)$ parametrized by Equation~(\ref{Eqn_CouplerParameterization}) has a sinusoidal oscillation period given by $T_{\lambda} = \pi \lambda_{\mathrm{deg}}/\mathrm{M}_{\sigma}$. To place these definitions in context, a directional coupler with an interaction length of $L=1 \,\, \mathrm{mm}$ and local linear coupling strength of $\kappa_{\sigma}(\lambda) = 1.053871 \times 10^{10} \lambda - 9217 \,\, \mathrm{m}^{-1}$ in the vicinity of $\lambda_{\mathrm{deg}} = 1550 \,\, \mathrm{nm}$ corresponds to the dimensionless parameters $\Delta\xi_{\sigma}=0.0494$ and $\mathrm{M}_{\sigma} = 16.335$ with a degeneracy splitting ratio of $\eta_{\sigma}(\lambda_{\mathrm{deg}}) = 1 - \sin^{2}\big(\xi_{\sigma}(\lambda_{\mathrm{deg}} ; \lambda_{\mathrm{deg}} )\big) = 0.4507$ and oscillation period of $T_{\lambda} = 298 \,\, \mathrm{nm}$. The properties of the photon pair state can likewise be generalized by defining a dimensionless non-degeneracy $\Lambda = \vert \lambda_{2,0} - \lambda_{1,0} \vert/\lambda_{\mathrm{deg}}$ as well as dimensionless photon bandwidths $\Delta\lambda / \lambda_{\mathrm{deg}}$, with $\Delta\lambda $ given in terms of full-width-half-maximum intensity.

\par The total parameter space of the near-degeneracy regime is then generated by the following variables: (i) $\Delta\xi _{\sigma}$, the dimensionless coupling offset, which can account for errors in design and fabrication or intentional detunings of the 50:50 split point from $\lambda_{\mathrm{deg}}$; (ii) $\mathrm{M}_{\sigma}\Lambda$, the product of the dimensionless first-order coupler dispersion with the dimensionless photon non-degeneracy, which gives the absolute difference in the coupling ($L\kappa_{\sigma}$) experienced at the two photon central wavelengths and is independent of $\lambda_{\mathrm{deg}}$; (iii) $\mathrm{M}_{\sigma}\Delta\lambda / \lambda_{\mathrm{deg}}$, the product of the dimensionless first-order coupler dispersion with the dimensionless photon bandwidths, which gives the absolute difference in coupling ($L\kappa_{\sigma}$) over the bandwidth interval $\Delta\lambda$, and is also independent of $\lambda_{\mathrm{deg}}$; (iv) the polarizations, $\alpha$ and $\beta$, of the twin photons.

\par The following subsections will discuss how each of these variables impacts IFPS. Polarization subscripts will be dropped in cases where only a single polarization is considered. For brevity, the substitution $\eta_{\sigma}(\lambda_{j,0}) \rightarrow \eta_{\sigma}^{(j)}$ ($j \in \{1,2\}$) will be made when referring to the splitting ratio at the photon central wavelengths.

\begin{figure}
	\includegraphics[width=1\columnwidth]{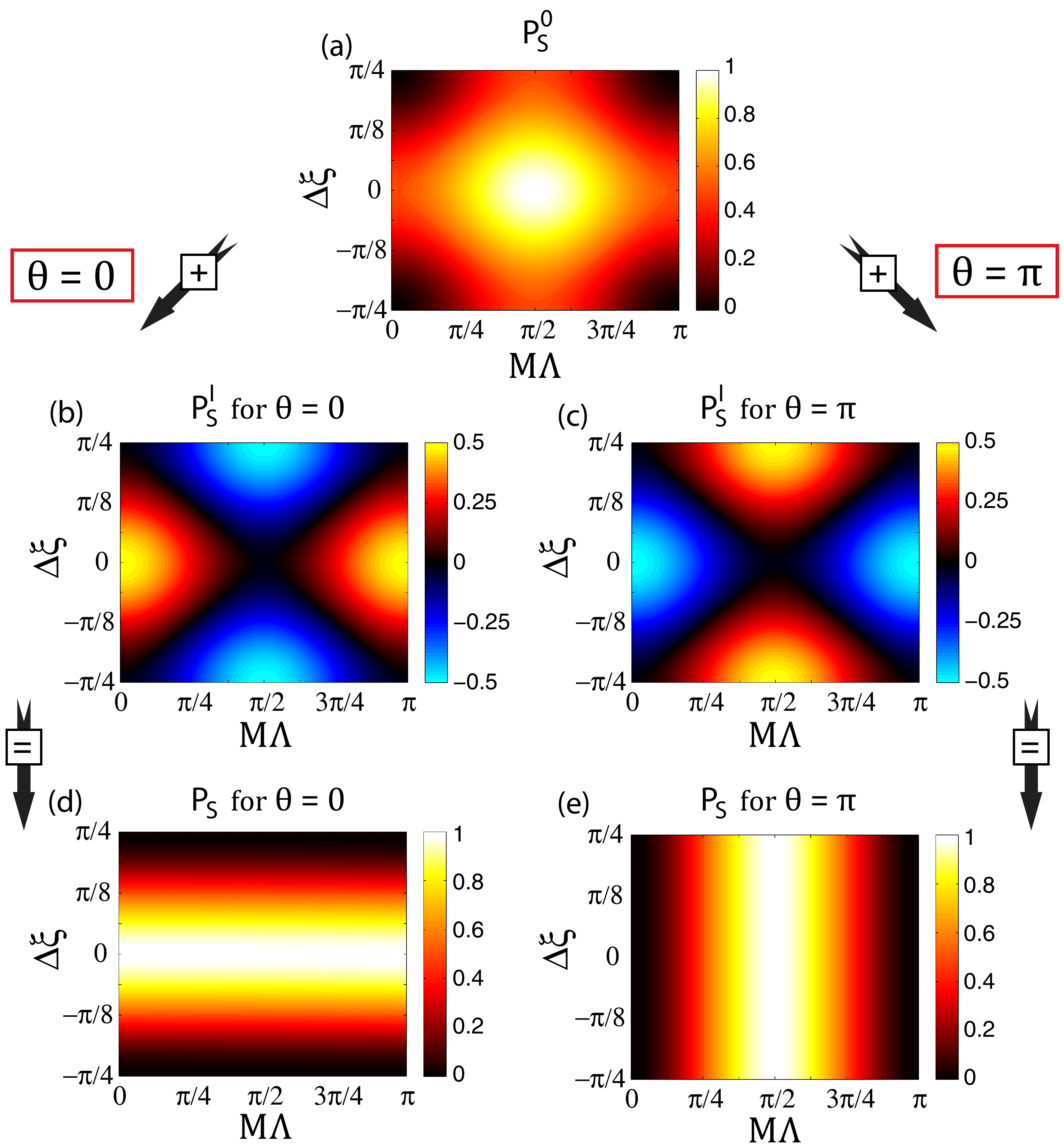}
	\caption{Separation probability $\left( P_{\textrm{S}} = P_{\textrm{S}}^{0} + P_{\textrm{S}}^{I} \right)$ as a function of coupler attributes in the near-degeneracy regime. The JSA used to model the photon state was based on Type I SPDC from a nonlinear waveguide, with further details given in the Appendix. The twin photons and nonlinear process pump were assigned dimensionless bandwidths of  $\Delta\lambda_{1}/\lambda_{\textrm{deg}}=\Delta\lambda_{2}/\lambda_{\textrm{deg}}= 3.205 \times 10^{-4}$ and $\Delta\lambda_{P}/\lambda_{\textrm{deg}}= 1.282 \times 10^{-4}$ (e.g. corresponding to $\Delta\lambda_{1(2)}=0.25 \, \mathrm{nm}$ and $\Delta\lambda_{P}= 0.1 \, \mathrm{nm}$ at $\lambda_{\mathrm{deg}} = 780 \, \mathrm{nm}$) respectively.} 
	 \label{Fig:CoPol_ContourPlots_1}
\end{figure}

\subsubsection{Salient Features}
\label{SubSubSec:SalientFeatures}

\par To highlight the main consequences of coupler dispersion in the near-degeneracy regime, the anti-bunched outcome probability $P_{\mathrm{S}}$, as well as its classical and non-classical components $P_{\mathrm{S}}^{0}$ and $P_{\mathrm{S}}^{I}$, were computed for a narrowband co-polarized state as a function of the coupling offset $\Delta\xi $ and dimensionless product $\mathrm{M} \Lambda$. Figures~\ref{Fig:CoPol_ContourPlots_1}(a)-(e) show contour plots of the calculation results at $\theta=0$ and $\theta=\pi$. The plotted span $\Delta\xi \in [-\pi/4, \pi/4]$ maps directly to degeneracy splitting ratios $\eta(\lambda_{\mathrm{deg}}) \in [0,1]$, and the span $\mathrm{M}\Lambda \in [0, \pi]$ covers all allowed permutations of the splitting ratio difference $\vert \eta^{(2)} - \eta^{(1)} \vert \in [0,1]$ for each $\Delta \xi$. All of the plotted functions are periodic and continuous beyond these axis limits.

\par
\begin{figure} [b!]
\centering
\includegraphics[width=0.8\columnwidth]{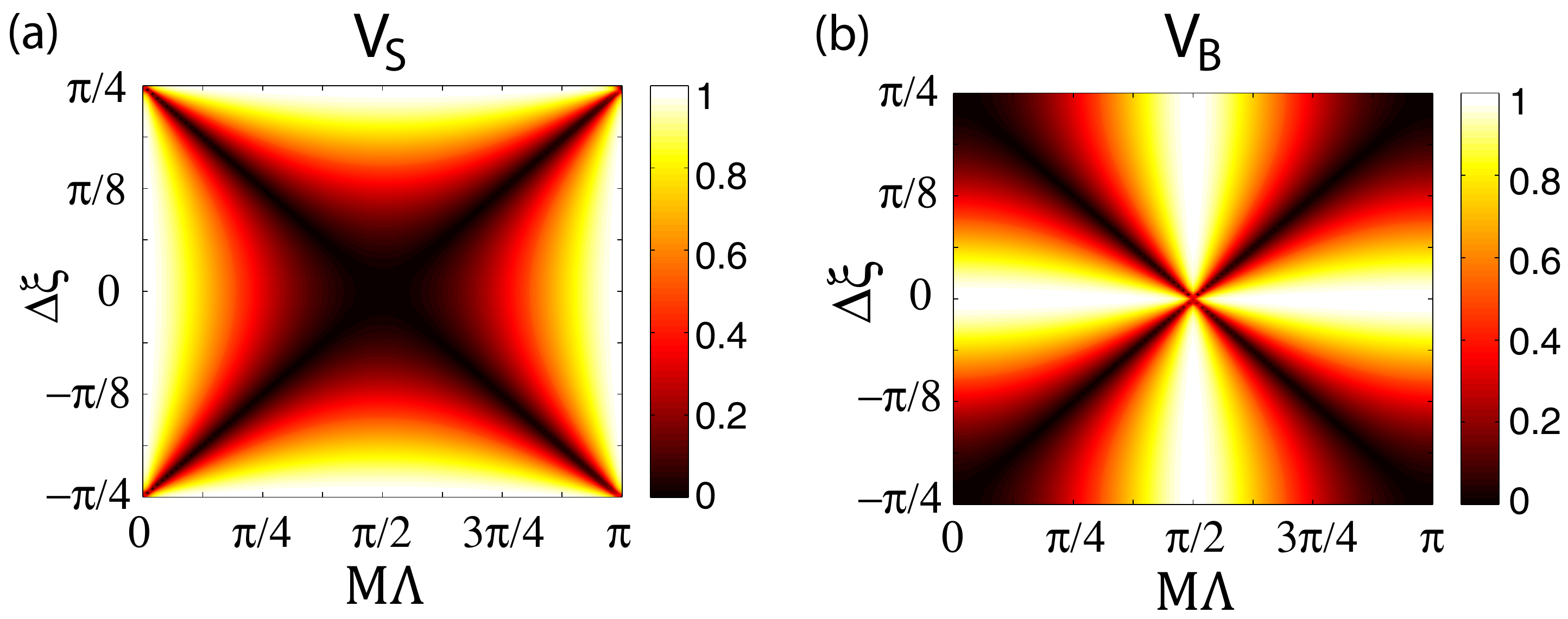}
\caption{Interference visibilities as a function of coupler attributes in the near-degeneracy regime; calculated for the same conditions as Fig.\ref{Fig:CoPol_ContourPlots_1}. A singularity exists in $V_{\textrm{B}}$ at the coordinate $\left(\Delta\xi =0, \mathrm{M}\Lambda = \pi/2 \right)$.} 
\label{Fig:CoPol_ContourPlots_2}
\end{figure}

 The classical separation probability $P_{\textrm{S}}^{0}$ is independent of the relative phase $\theta$. However, a $\pi$ phase shift in $\theta$ has the effect of inverting the sign of $P_{\textrm{S}}^{I}$, which in turn influences the total separation probability $P_{\textrm{S}}$. Note that the $P_{\textrm{S}}^{I}$ plots have regions that undergo a sign change similar to, but independent of, an applied $\pi$ phase shift. This occurs due to a change of sign in the mode operator transformation matrix elements [Eqn.~(\ref{Eqn_ModeTransformation})]. The sign change occurs smoothly without discontinuity, along a boundary where $P_{\textrm{S}}^{I}= 0$. Figure~\ref{Fig:CoPol_ContourPlots_2} shows the corresponding interference visibilities calculated from $P_{\textrm{S}}^{0}$, $P_{\textrm{B}}^{0} = 1 - P_{\textrm{S}}^{0}$, and $\left\vert P_{\textrm{S}}^{I} \right\vert = \left\vert P_{\textrm{B}}^{I} \right\vert$. The features of Figures~\ref{Fig:CoPol_ContourPlots_1}~and~\ref{Fig:CoPol_ContourPlots_2} will be discussed together.

\par The line $\mathrm{M}\Lambda =0$ corresponds to the evolution of any state under conditions of vanishing coupler dispersion ($\mathrm{M}=0$), and is therefore the portion of the parameter space accessible to bulk-optics beamsplitters. This line also describes the behaviour seen in any experiment utilizing an integrated coupler that is performed at degeneracy ($\Lambda=0$) even when the coupler dispersion is non-trivial. The behaviour here is well-known: away from $\Delta\xi = 0$ the bunched-state amplitudes acquire unequal weightings, and $P_{\mathrm{S}}$ degrades through imperfect amplitude cancellation. The dependence of $V_{\mathrm{B}}$ on the splitting ratio (calculated from $\Delta\xi$) agrees with Equation~(\ref{Eqn:IFPS_VB}), while $V_{\mathrm{S}}$ remains equal to unity as expected.

\par When the product $\mathrm{M}\Lambda$ is no longer vanishing, i.e. both coupler dispersion and non-degeneracy are present, several interesting and previously unstudied behaviours are revealed. Most prominently, for $\theta = 0$ and a coupling offset of $\Delta\xi = 0$ (perfect 50:50 splitting at the degeneracy wavelength $\lambda_{\mathrm{deg}}$), the anti-bunched outcome probability $P_{\mathrm{S}}$ in Fig.~\ref{Fig:CoPol_ContourPlots_1}(d) appears unaffected by coupler dispersion for any level of non-degeneracy, giving $P_{\mathrm{S}}=1$ for all $\mathrm{M}\Lambda$. Indeed, so long as the assumptions of the near-degeneracy regime remain valid, the only prerequisites for perfect deterministic separation aside from source indistinguishability are that $\theta = 0$, $\Delta\xi = 0$ and that the photons are co-polarized. This seems remarkable because perfect performance is maintained even though the splitting ratios $\eta^{(1)}$ and $\eta^{(2)}$ at the photon central wavelengths deviate from 50:50 as well as from one-another. The visibility $V_{\textrm{B}}$ also remains equal to unity along this line, except for a singularity at $\left(\Delta\xi =0, \mathrm{M}\Lambda = \pi/2 \right)$.

\par
\begin{figure} [h]
\centering
\includegraphics[width=0.8\columnwidth]{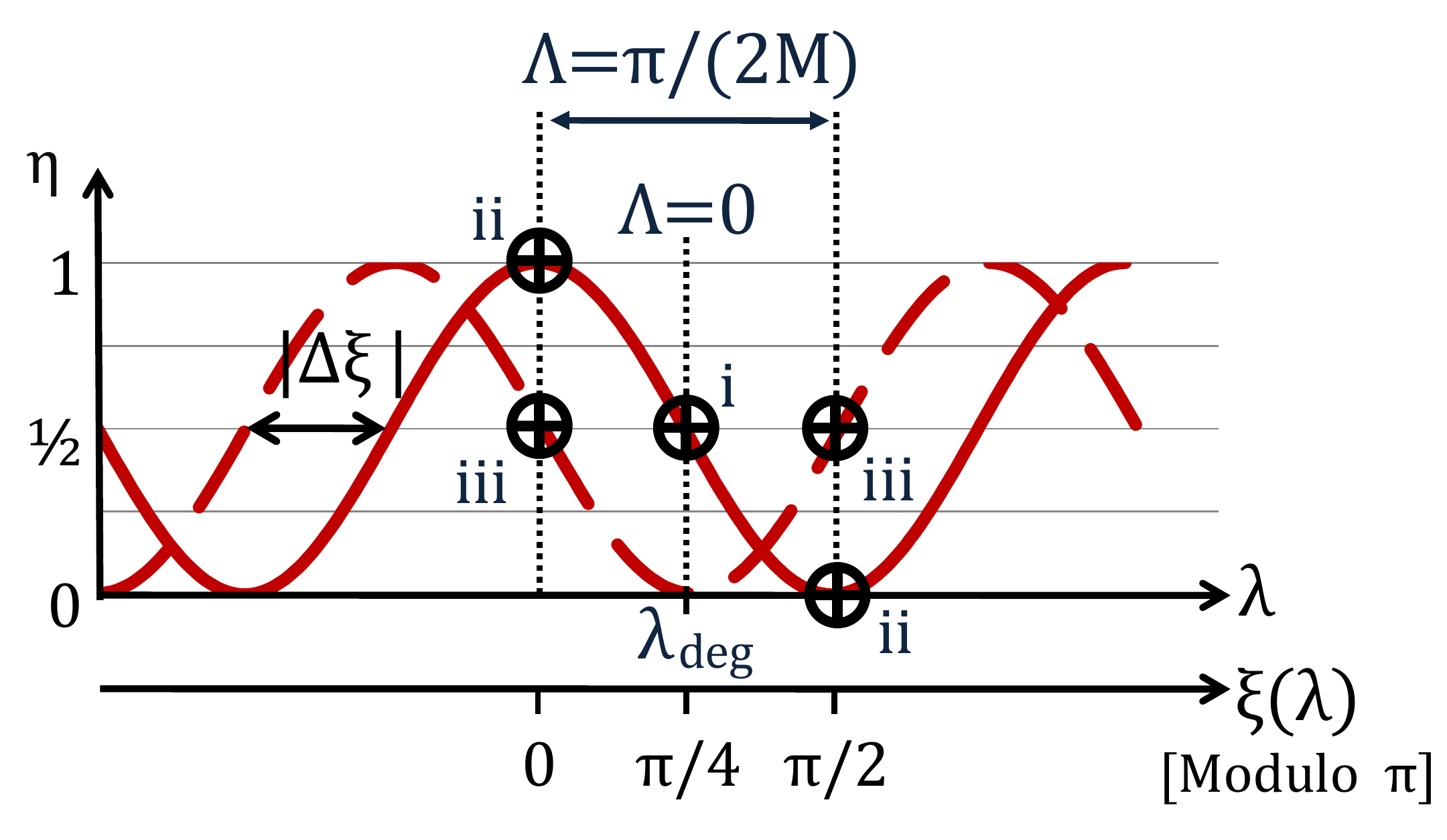}
\caption{Visualization of the interplay between $\Delta\xi$ and $\mathrm{M}\Lambda$ in determining $\eta^{(1)}$ and $\eta^{(2)}$, showing operating points where the coupler responds to the twin-photon state as a perfect 50:50 splitter (i and iii) and a perfect wavelength-demultiplexer (ii).} 
\label{Fig:AntisymmetryIllustration}
\end{figure}

 To understand why $P_{\mathrm{S}}$ in Fig.~\ref{Fig:CoPol_ContourPlots_1}(d) remains unity along the line $\Delta\xi=0$, first consider how the coupler responds at each $\mathrm{M}\Lambda$ in terms of the power splitting ratio $\eta(\lambda)$. At the two extremes, $\mathrm{M}\Lambda = 0$ and $\mathrm{M}\Lambda = \pi$, the splitting ratio satisfies $\eta^{(1)} = \eta^{(2)} = 1/2$ so that the coupler behaves as an ideal 50:50 splitter. At the central coordinate, $\mathrm{M}\Lambda = \pi/2$, the coupler instead behaves as a perfect wavelength demultiplexer (WD) with $\big\vert \eta^{(2)}- \eta^{(1)} \big\vert = 1$, and deterministically separates the photons classically without interference. Within the intermediate regime between $\mathrm{M}\Lambda = 0$ and $\mathrm{M}\Lambda = \pi/2$, the coupler transitions from a perfect 50:50 splitter to a perfect WD in such a manner that reduction in $P^{I}_{\mathrm{S}}$ due to the loss of interference is compensated for by increases in $P^{0}_{\mathrm{S}}$ from WD-induced splitting, as illustrated in Figure~\ref{Fig:AntisymmetryIllustration}.

\par A few other comments about the WD behaviour at $\left(\Delta\xi =0, \mathrm{M}\Lambda = \pi/2 \right)$ should be made. The interference visibilities at this coordinate, in accordance with Equation~(\ref{Eqn_InterferenceVisibility}), become $V_{\mathrm{S}}=0$ and $V_{\mathrm{B}} = \lim_{(x,y \to 0)} \frac{x}{y}$. The former vanishes because the separation  is entirely classical; the latter becomes undefined due to a combination of this same reason with a vanishing of the classical bunched outcome probability $P_{\textrm{B}}^{0}= 1 - P_{\textrm{S}}^{0}$. Also, although WD behaviour deterministically separates the photons based on wavelength, the frequency degree of freedom nonetheless remains uncorrelated with the output port because of the path superposition at the input, e.g. an input state $\vert \psi \rangle = \vert \lambda_{1,0} \rangle_{A} \vert \lambda_{2,0} \rangle_{A} +  \vert \lambda_{1,0} \rangle_{B} \vert \lambda_{2,0} \rangle_{B}$ maps to the output state $\vert \psi \rangle = \vert \lambda_{1,0} \rangle_{A} \vert \lambda_{2,0} \rangle_{B} +  \vert \lambda_{1,0} \rangle_{B} \vert \lambda_{2,0} \rangle_{A}$. 

\par Now consider Figure~\ref{Fig:CoPol_ContourPlots_1}(e), where $\theta=\pi$. Interestingly, there is a level of coupler dispersion at which error in the degeneracy 50:50 split point no longer impacts the separation fidelity. This is seen in along the line $\mathrm{M}\Lambda = \pi/2$, where $P_{\mathrm{S}}=1$ at any $\Delta\xi$, which is again attributable to splitting ratio anti-symmetry.

\par The behaviour of the anti-bunched interference visibility $V_{\mathrm{S}}$ is also of interest. Dispersion leads to the novel situation of having an interference visibility that can change while the outcome probability $P_{\textrm{S}}$ remains the same. Changes to $V_{\mathrm{S}}$ are entirely dispersion-driven and in response to dissimilarities in $\eta^{(1)}$ and $\eta^{(2)}$. This has no parallel in bulk optics. Notably, $V_{\mathrm{S}}$ also gives a useful indication of the extent to which the coupler is responding as a beamsplitter or a WD.

\FloatBarrier 
\subsubsection{Photon Polarization Diversity}
\label{SubSubSec:PolarizationDiversity}

\par
\begin{figure} [h!]
\centering
\includegraphics[width=0.7\columnwidth]{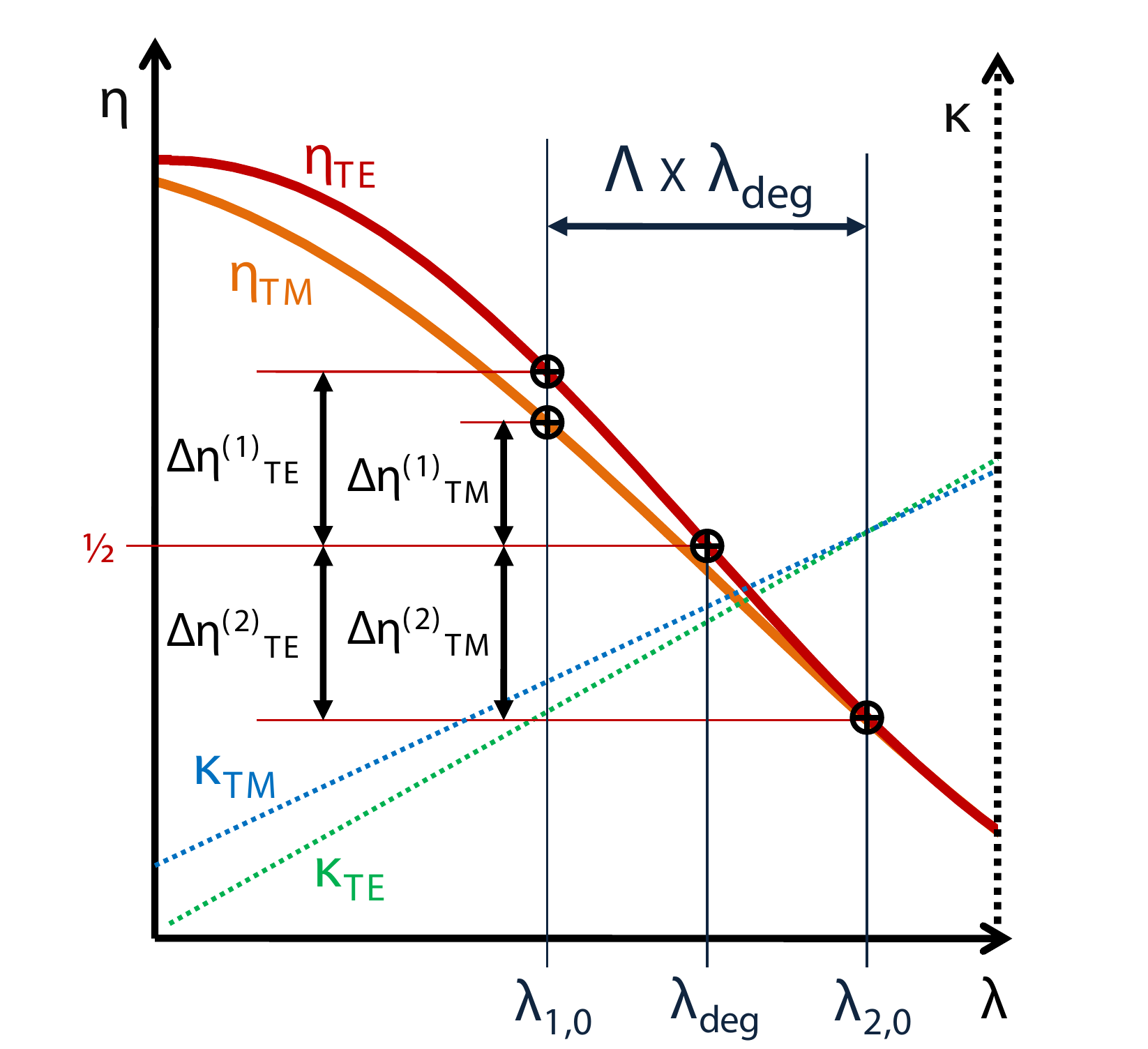}
\caption{Visual depiction of the antisymmetry conditions leading to $\eta^{(1)}+\eta^{(2)}=1$; as an example these are shown to be satisfied for a $\mathrm{TE}$-$\mathrm{TE}$ co-polarized state, but broken for other polarizations (e.g. $\eta^{(1)}_{\mathrm{TM}} \neq \eta^{(2)}_{\mathrm{TM}} \neq \eta^{(2)}_{\mathrm{TE}}$) due to coupler birefringence in $\eta_{\sigma}(\lambda_{\mathrm{deg}})$ and $\mathrm{M}_{\sigma}$; note that $\eta_{\mathrm{TE}}^{(2)} = \eta_{\mathrm{TM}}^{(2)}$ by coincidence only.} 
\label{Fig:PolarizationDiversityIllustration}
\end{figure}

 Many photon pair sources are capable of generating cross-polarized states. Several also possess the ability to generate both $\mathrm{TE}$-$\mathrm{TE}$ and $\mathrm{TM}$-$\mathrm{TM}$ co-polarized states concurrently \cite{Kang_OL_2012}. Introducing additional states of polarization creates two main complications. Firstly, polarization-dependent coupling strengths may lead to differences in the degeneracy splitting ratio, i.e. $\eta_{\mathrm{TE}}(\lambda_{\mathrm{deg}}) \neq \eta_{\mathrm{TM}}(\lambda_{\mathrm{deg}})$, as well as birefringence in the linear coupler dispersion, i.e. $\mathrm{M}_{\mathrm{TE}} \neq \mathrm{M}_{\mathrm{TM}}$. Both of these non-idealities are potential sources of asymmetry that cause deviations from $\eta^{(1)}+\eta^{(2)}=1$, as depicted in Figure~\ref{Fig:PolarizationDiversityIllustration}. Numerous strategies are available to engineer the polarization dependence for better performance.

\par The second complication arises when polarization entanglement is present but is not maximal. In this case two distinct tuning curves exist (e.g. see Fig. 2 in Ref.~\cite{Horn_SciRep_2013}), one for each polarization, and the degeneracy wavelength occurs at their intersection. Away from degeneracy, it becomes necessary to consider four different photon central wavelengths and two possible non-degeneracies, i.e. $\Lambda_{\mathrm{TE-TM}}$ and $\Lambda_{\mathrm{TM-TE}}$. This may restrict the available options for improving performance when faced with coupling birefringence. On-chip IFPS has not previously been demonstrated for more than a single polarization \cite{Silverstone_NaturePhotonics_2014, Jin_PRL_2014}.

\subsubsection{Photon Bandwidth Effects}
\label{SubSubSec:BandwidthEffects}

\par The discussion has thus-far concentrated on the photon central wavelengths, but bandwidth effects can be similarly investigated for a single polarization using the previously-introduced dimensionless parameter $\mathrm{M}\Delta\lambda/\lambda_{\mathrm{deg}}$. To put this parameter into physical context: the value $\mathrm{M}\Delta\lambda/\lambda_{\mathrm{deg}} = \pi$ occurs when the photon bandwidth spans one full oscillation cycle in $\eta(\lambda)$. In the calculations that follow, the coupler offset $\Delta\xi$ is zero, the photons are both of equal bandwidth, and the photon sources are based on an SPDC process as was used in the calculation of Figures~\ref{Fig:CoPol_ContourPlots_1} and \ref{Fig:CoPol_ContourPlots_2}.

\par
\begin{figure}[!t]
\includegraphics[width=1\columnwidth]{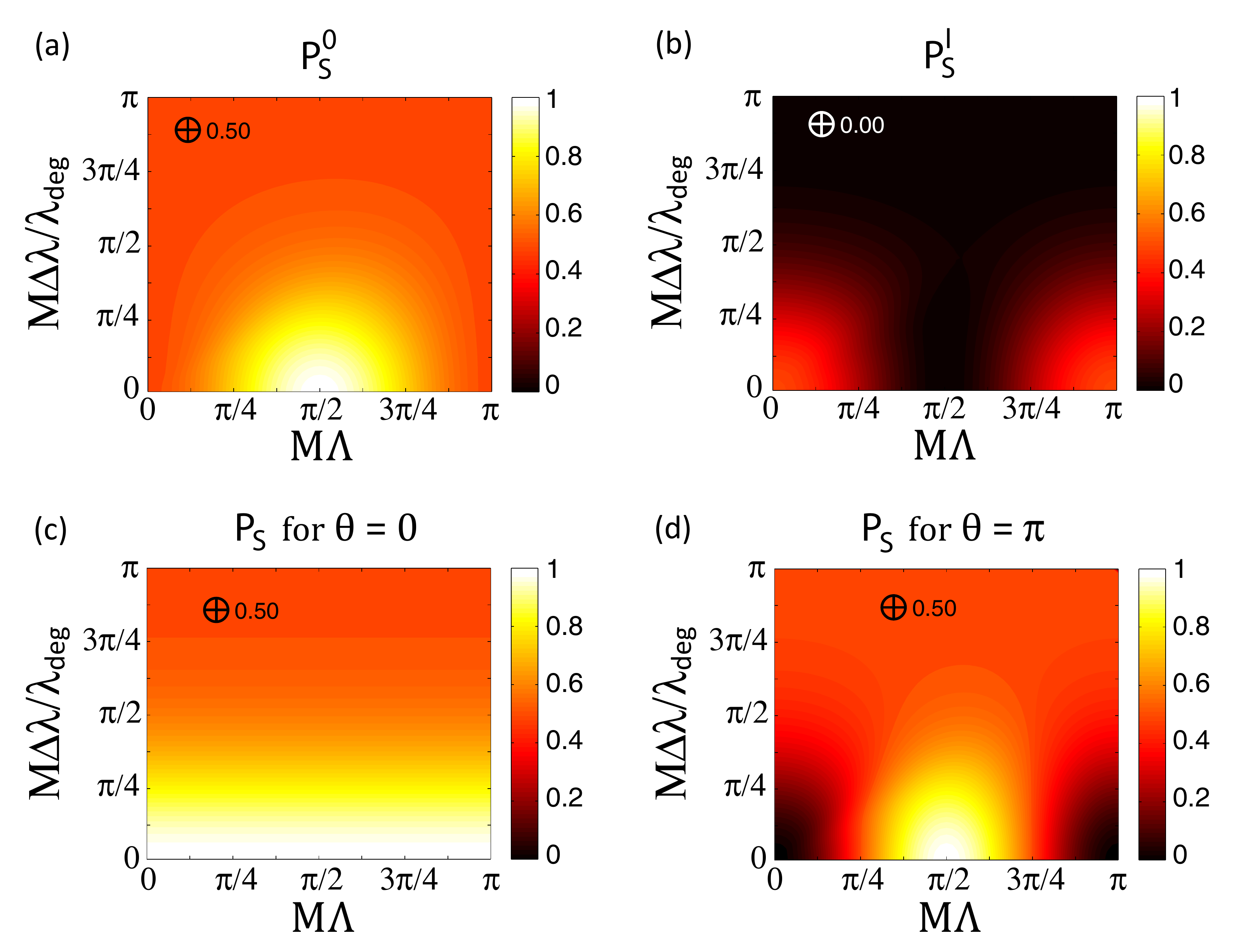}
\caption{Separation probability as a function of the dimensionless products $\mathrm{M}\Delta\lambda/\lambda_{\mathrm{deg}}$ and  $\mathrm{M}\Lambda$, calculated for spectrally-uncorrelated ($\textrm{SN}=1$) co-polarized photons via SPDC.} 
\label{Fig_BandwidthEffects}
\end{figure}

 Figure~\ref{Fig_BandwidthEffects} shows how $P_{\textrm{S}}$, $P_{\textrm{S}}^{0}$, and $P_{\textrm{S}}^{I}$ vary as a function of $\mathrm{M}\Delta\lambda/\lambda_{\mathrm{deg}}$ and  $\mathrm{M}\Lambda$, for a spectrally uncorrelated photon pair. The observed behaviour is in agreement with Figure~\ref{Fig:CoPol_ContourPlots_1} in the limit of vanishing bandwidth, i.e. $\mathrm{M}\Delta\lambda/\lambda_{\mathrm{deg}}\rightarrow 0$. However, as the dimensionless bandwidth-dispersion product increases towards $\mathrm{M}\Delta\lambda/\lambda_{\mathrm{deg}} = \pi$, the interference term vanishes and both $P_{\textrm{S}}$ and $P_{\textrm{S}}^{0}$ approach values of 0.5. To understand this, it is important to recall from Equations~(\ref{Eqn_R0})-(\ref{Eqn_RI}) that the probabilities are a result of an integration over many possible permutations of $\omega_{1}$ and $\omega_{2}$. The permutations that contribute are ultimately determined by the JSA, $\phi^{j}_{\alpha \beta}(\omega_{1}, \omega_{2})$. When $\mathrm{M}\Delta\lambda/\lambda_{\mathrm{deg}}$ increases, the contributing $\omega$ permutations begin to straddle both the beamsplitter and WD behaviour of the coupler; furthermore, many of these permutations do not conform to the ideal $\eta^{(1)}+\eta^{(2)} = 1$ splitting ratio anti-symmetry. This results in a `washing-out' of the interference. For frequency pairings in the integration that do contribute a positive value towards $P_{\textrm{S}}^{I}$, there are also pairings where the mode operator transformation undergoes a sign inversion (see Section~\ref{Sec:Background}) that negates these contributions. The classical WD enhancement at $\mathrm{M}\Lambda = \pi/2$ also disappears as $\mathrm{M}\Delta\lambda/\lambda_{\mathrm{deg}}$ increases, because a shrinking percentage of contributing $\omega_{1}$ and $\omega_{2}$ permutations satisfy the demultiplexing requirement $\left\vert \eta^{(2)} - \eta^{(1)} \right\vert = 1$.

\par
\begin{figure}[b!]
\centering
\includegraphics[width=0.8\columnwidth]{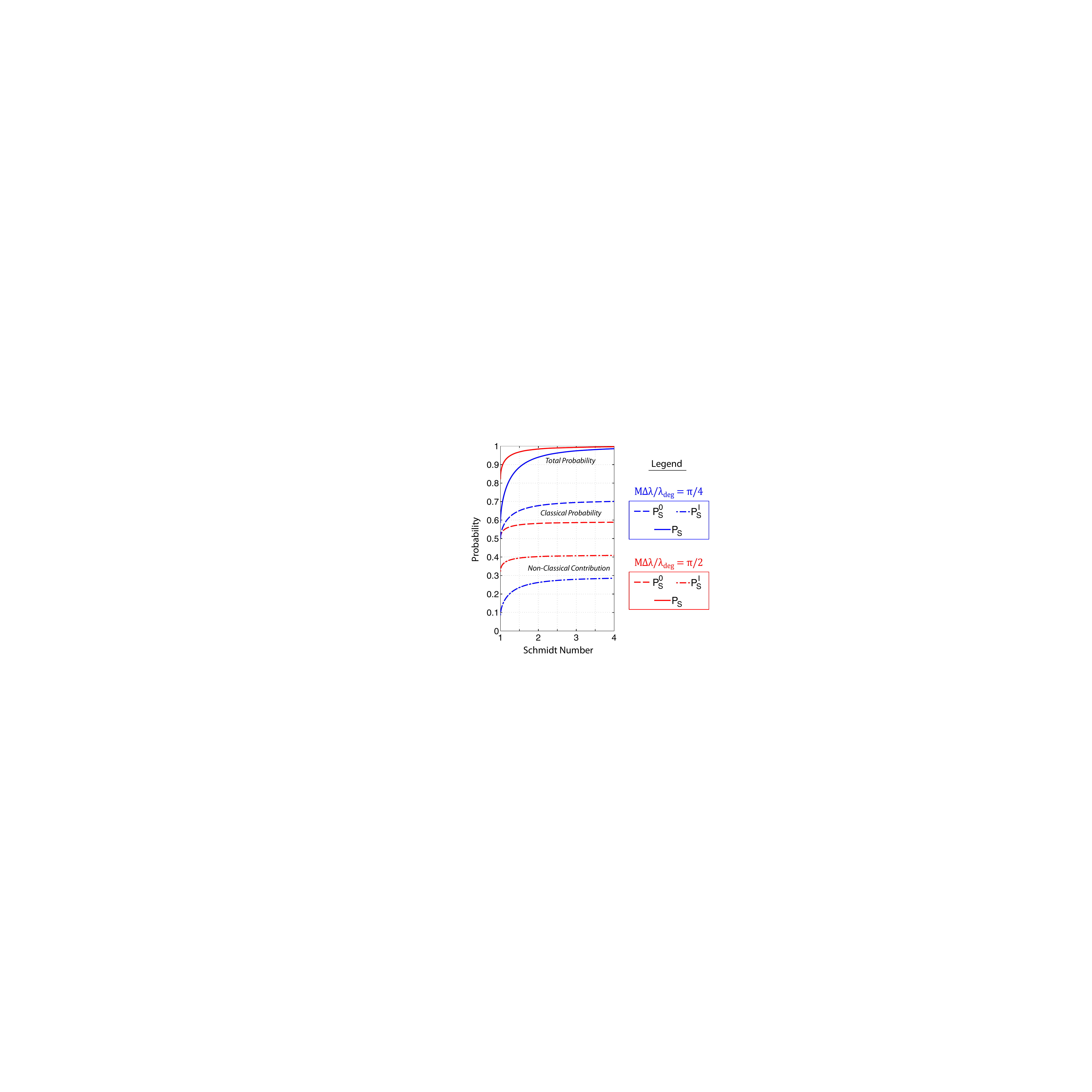}
\caption{Calculated $P_{\textrm{S}}$ versus Schmidt Number for co-polarized degenerate photons, shown at two levels of $\mathrm{M}\Delta\lambda/\lambda_{\mathrm{deg}}$. The state JSA was based on SPDC and entanglement was varied by narrowing the pump bandwidth for a fixed photon pair bandwidth.} 
\label{Fig:BW_SN_Dependence}
\end{figure}

 It turns out that spectral entanglement can influence the bandwidth-dependence of IFPS by restricting the $\omega_{1}$,$\omega_{2}$ combinations that can contribute towards $P_{\textrm{S}}^{0}$ and $P_{\textrm{S}}^{I}$. This effect is shown in Figure~\ref{Fig:BW_SN_Dependence}; as the state's spectral entanglement and hence Schmidt Number is increased, $P_{\textrm{S}}$ is asymptotically restored towards unity through increases in $P_{\textrm{S}}^{0}$ and $P_{\textrm{S}}^{I}$. In SPDC, spectral entanglement depends partly on the dispersion properties of the source \cite{Eckstein_PRL_2011, Kang_JOSAB_2014}. However, it is generally dominated by the amount of wavelength anti-correlation imparted by the pump bandwidth. Narrower pump bandwidths lead to greater anti-correlation and thus enhanced entanglement. For SPDC, the increased frequency anti-correlation means that the JSA $\phi^{j}_{\alpha \beta}(\omega_{1}, \omega_{2})$ tends to be non-vanishing only for frequency permutations where $\omega_{1} + \omega_{2} \approx \omega_{\textrm{p},0}$, with $\omega_{\textrm{p},0}$ being the pump central wavelength. In SFWM, this constraint involves an additional pump term on the righthand side. In either case, the result is that the non-vanishing frequency permutations tend to be equidistant from the degeneracy point. This has the side effect of enforcing (albeit imperfectly) the ideal $\eta^{(1)} + \eta^{(2)} = 1$ antisymmetry condition that maximizes $P_{\mathrm{S}}$, hence why entanglement tends to restore the separation performance towards $P_{\mathrm{S}}=1$.

\par It is insightful to consider to what extent these photon bandwidth effects pose a practical limit to IFPS. The above analysis has shown that degradation to the separation probability is most severe for spectrally uncorrelated states. However, to place Figure~\ref{Fig_BandwidthEffects} in perspective, note that $P_{\mathrm{S}}$ remains above 90\% for $\mathrm{M}\Delta\lambda/\lambda_{\mathrm{deg}} < \pi/8$, and only drops below 80\% when $\mathrm{M}\Delta\lambda/\lambda_{\mathrm{deg}}$ exceeds $\pi/4$. To reach $\mathrm{M}\Delta\lambda/\lambda_{\mathrm{deg}} = \pi/4$, a splitting-ratio oscillation period of $T_{\lambda} = 4\Delta\lambda$ or smaller is required. For a photon bandwidth of $\Delta\lambda = 3 \,\, \mathrm{nm}$ at the common degeneracy point of $\lambda_{\mathrm{deg}} = 1550 \,\, \mathrm{nm}$, this requires a dimensionless dispersion of $\mathrm{M}= 405.8$ and a corresponding oscillation period of $T_{\lambda} = 12 \,\, \mathrm{nm}$. Serious degradation in $P_{\mathrm{S}}$ at bandwidths on the order of a few nanometers is therefore unlikely except in cases of severe coupler dispersion. 

\subsubsection{Numerical Example}
\label{SubSubSec:NumericalExample}

\par The following example shows several of the behaviours described above, for $\Delta\xi = 0$ and $\theta = 0$. IFPS performance was computed for the silicon-on-silica coupler described in Section~\ref{Sec:Background}. Its coupling strength in the vicinity of $\lambda_{\textrm{deg}}=\textrm{780~nm}$ is approximately linear, with $\mathrm{M} = 4.072$. Table~\ref{Table:Example_SummaryTable} shows the results for several different two-photon states. Note that the non-degeneracies of 100~nm and 200~nm map approximately to $\mathrm{M}\Lambda =\pi/6 $ and $\mathrm{M}\Lambda =\pi/3 $ respectively. Similarly, the bandwidths 10~nm and 100~nm map to $\mathrm{M}\Delta\lambda/\lambda_{\textrm{deg}}$ values of $\sim$ $\pi/60$ and $\pi/6$.

\begin{table}[ht!]
\caption{IFPS Performance for Silica-on-Silicon Coupler}
\centering 
\begin{tabular}{c c c || c  c } 
\hline 
$\vert \lambda_{2,0} - \lambda_{1,0} \vert$ & $\Delta\lambda$   & SN  & $P_{\textrm{S}}$ & $V_{\textrm{S}}$ \\ [0.5ex] 
\hline\hline \\ [-3ex]
0~nm      & 10~nm     & 1.00       & 0.999     & 0.998 \\ [0ex] 
0~nm      & 100~nm   & 1.00       &  0.915    & 0.827 \\ [0ex] 
0~nm      & 100~nm   & 1.26       &  0.976    & 0.833 \\ [0ex] 
100~nm  & 10~nm     & 1.00       & 0.997     & 0.602 \\ [0ex] 
200~nm  & 10~nm     & 1.00       &  0.966    & 0.130 \\ [0ex] 
\hline
\end{tabular}
\label{Table:Example_SummaryTable}
\end{table}

\par Comparison of the first and second rows shows the decrease in splitting fidelity when the photon bandwidth is large. Row three shows how this is mitigated in the presence of spectral entanglement. In row four, the non-degeneracy is increased to 100~nm, yet the coupler continues to perform well. However, row five shows that this performance begins to degrade at 200~nm. This is because the coupling strength for this device, while locally linear near 780~nm, becomes a polynomial in $\lambda$ at large non-degeneracies, leading to splitting ratio asymmetry and degradation in IFPS performance.  Such asymmetries are discussed in Section~\ref{SubSec:DesigningForLargeNonDeg}.

\FloatBarrier 
\subsection{Designing for Large Tunable Non-Degeneracies}
\label{SubSec:DesigningForLargeNonDeg}

\par Given the emergence of highly-tunable integrated sources, where the non-degeneracy can be tuned from zero to several hundred nanometers \cite{Horn_SciRep_2013}, it is important to consider IFPS performance far from degeneracy, and far beyond what has been previously demonstrated for on-chip interference.

\subsubsection{Extrapolating the Near-Degeneracy Behaviour}

\par In the near-degeneracy regime, antisymmetry in the splitting ratio deviations from 50:50 allowed the anti-bunched outcome probability $P_{\mathrm{S}}$ to remain independent of the dimensionless non-degeneracy $\Lambda$ for $\Delta\xi=0$ and $\theta = 0$. As seen in the example of Section~\ref{SubSubSec:NumericalExample}, this can also be a good approximation of behaviour for non-degeneracies of up to several hundred nm, provided the coupling strength is sufficiently linear. Judicious choice of the material system and waveguide design aimed at minimizing coupler dispersion can help maximize the range of $\Lambda$ for which the assumptions of the near-degeneracy regime are valid and perfect IFPS fidelity persists.

\begin{figure}[h]
\centering
\includegraphics[width=1\columnwidth]{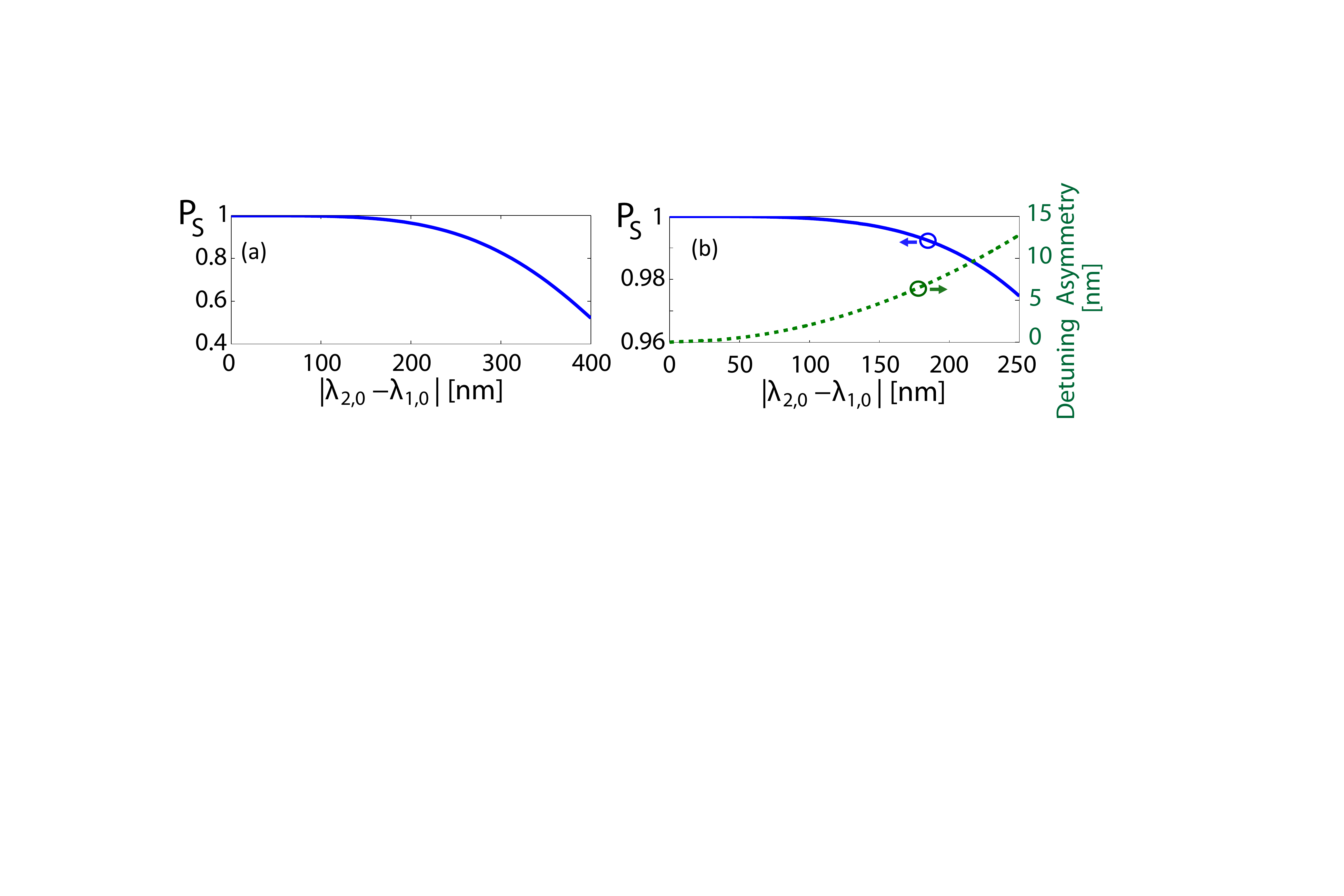}
\caption{Loss of IFPS separation fidelity due to: (a) coupling strength non-linearity, based on the coupler example in Sec.~\ref{SubSubSec:NumericalExample}; and (b) tuning curve asymmetry, computed for an ideal linear coupling strength, using the SPDC tuning curve of the device reported in Ref.~\cite{Horn_SciRep_2013}).} 
\label{Fig:LossOfSymmetry}
\end{figure}

\par The performance near-degeneracy eventually breaks down, leading to a loss of separation fidelity as seen in Figure~\ref{Fig:LossOfSymmetry}. There are two main reasons for this. Firstly, the local linearity of $\kappa(\lambda)$ may inaccurately describe the coupling strength's global behaviour, with higher-order coupler dispersion causing chirp in the splitting ratio oscillations. Secondly, as the central wavelengths $\lambda_{1,0}$ and $\lambda_{2,0}$ further separate and follow the phase-matching tuning curve of the nonlinear process, they eventually become asymmetric about the degeneracy wavelength. Together, these non-idealities lead to loss of antisymmetry of the central wavelength splitting ratios with respect to their degeneracy value $\eta(\lambda_{\mathrm{deg}})$.

\subsubsection{Loss of Splitting Ratio Antisymmetry}

\par Asymmetry negates the classical WD compensation to $P_{\mathrm{S}}$ by causing deviations from the ideal $\eta^{(1)}+\eta^{(2)}=1$ condition. Additionally, it does not allow for the construction of convenient dimensionless variables that span all parameter space, as was done in Section~\ref{SubSec:NearDegeneracy}. The details of how $P_{\mathrm{S}}$ and the interference visibilities $V_{\mathrm{B}}$ and $V_{\mathrm{S}}$ behave as a function of the photon bandwidths and non-degeneracies therefore become case-specific to the state and coupler properties involved. Specific cases remain calculable from the equations of Section~\ref{SubSec:General_IFPS_Expressions}. Nonetheless, generic comments about performance can still be made. For example, if the IFPS behaviour for a given state and coupler were plotted in terms of the variables $\eta(\lambda_{\textrm{deg}})$ and $\left\vert \lambda_{2,0} - \lambda_{1,0}\right\vert$ instead of $\Delta\xi$ and $\mathrm{M}\Lambda$ respectively, the global behaviour is generally similar to that of Figure~\ref{Fig:CoPol_ContourPlots_1} except that it would deviate from the $\left\vert \lambda_{2,0} - \lambda_{1,0}\right\vert$ axis as the non-degeneracy is increased. 

\begin{figure}[t]
\centering
\includegraphics[width=0.8\columnwidth]{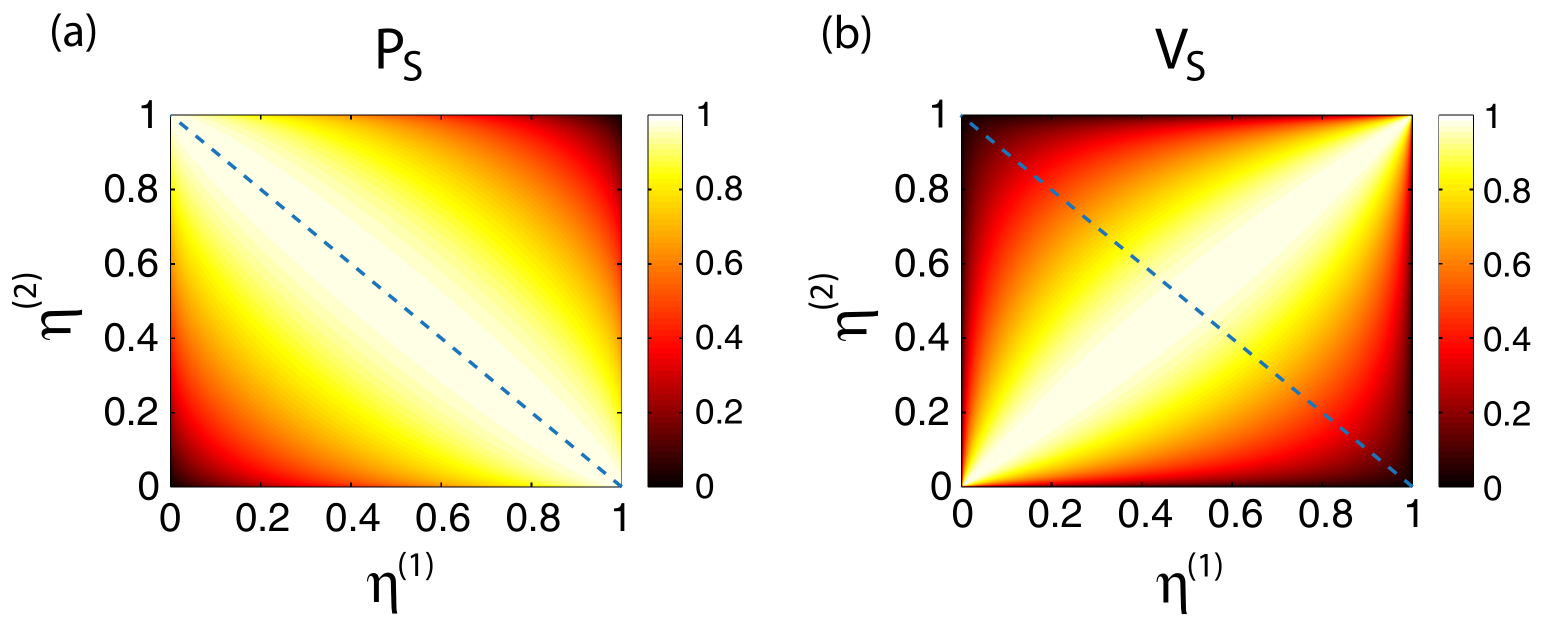}
\caption{Anti-bunched outcome probability $P_{\mathrm{S}}$ and interference visibility $V_{\mathrm{S}}$ at $\theta=0$ for all permutations of the photon central wavelength splitting ratios $\eta^{(1)}$ and $\eta^{(2)}$; dashed lines indicate the contour $\eta^{(1)} + \eta^{(2)} = 1$ along which $P_{\mathrm{S}}$ attains its maximum value.}
\label{Fig_ContourPlots_Global}
\end{figure}

\par
\begin{figure}[t!]
\centering
\includegraphics[width=0.8\columnwidth]{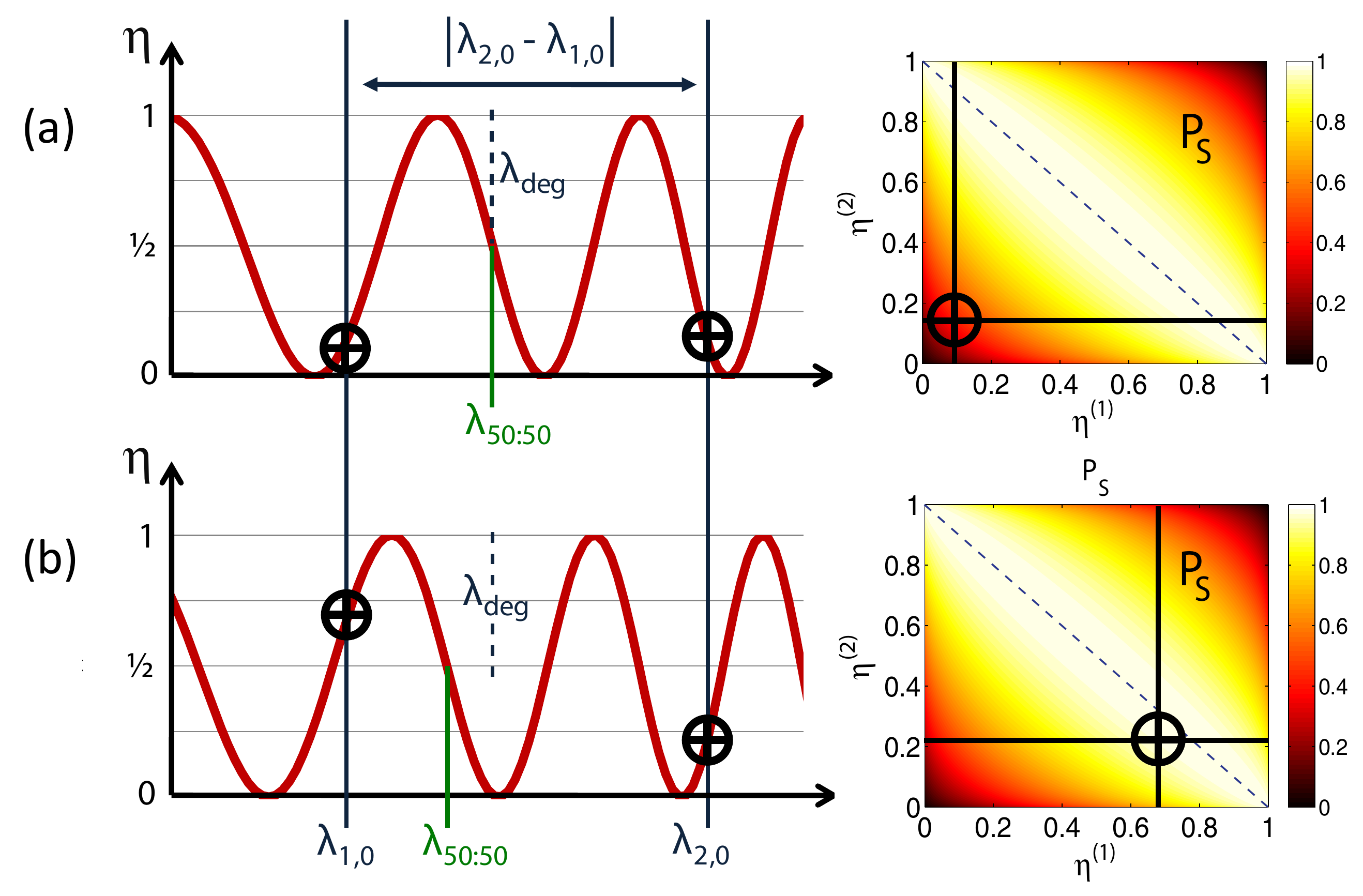}
\caption{(a) Non-ideal operating conditions at an arbitrary central wavelength separation; $\eta$ is chirped by a quadratic dependence of $\kappa$ on $\lambda$, with $\lambda_{1,0}$ and $\lambda_{2,0}$ asymmetric about degeneracy. (b) Performance can be restored through adjustment of $\lambda_{\mathrm{50:50}}$ away from the degeneracy point.}
\label{Fig_TunableDesignPoint}
\end{figure}

 When faced with a breakdown of the near-degeneracy assumptions, it becomes useful to consider general performance in terms of the central wavelength splitting ratios $\eta^{(1)}$ and $\eta^{(2)}$. Figure~\ref{Fig_ContourPlots_Global} depicts $P_{\mathrm{S}}$ and $V_{\mathrm{S}}$ for all possible permutations of these parameters; the behaviour of $V_{\mathrm{B}}$ is merely the left-to-right mirror image of $V_{\mathrm{S}}$. High separation fidelities ($P_{\mathrm{S}} > 0.9$) remain achievable so long as $\eta^{(1)}$ and $\eta^{(2)}$ are on opposing sides of the 50:50 split point, corresponding to the top-left and bottom-right quadrants of Figure~\ref{Fig_ContourPlots_Global}(a). This can generally be satisfied when $\lambda_{1,0}$ and $\lambda_{2,0}$ span no more than a single cycle in $\eta(\lambda)$, provided $\eta(\lambda_{\mathrm{deg}}) \approx 1/2$. Beyond one cycle it is possible to find $\eta^{(1)}$ and $\eta^{(2)}$ either both above or both below $\eta = 1/2$ such as in Figure~\ref{Fig_TunableDesignPoint}(a), with a net effect similar to a $\Lambda$-dependent non-zero $\Delta \xi$ in the near-degeneracy regime. As discussed earlier, implementations that minimize coupler dispersion can mitigate these non-idealities. 

\par In some cases, active compensation of the coupler performance may be desirable. One route to achieving this is illustrated in Figure~\ref{Fig_TunableDesignPoint}(b), where a translation in the 50:50 splitting wavelength $\lambda_{\textrm{50:50}}$, akin to introducing $\Delta\xi \neq 0$, shifts operation towards the ideal $\eta^{(1)} + \eta^{(2)} = 1$ conditions.

\section{Discussion}
\label{Sec:Discussion}

\par Directional coupler dispersion, as was shown in Section~\ref{Sec:IFPS_ImpactOfCouplerAttributes}, can have a significant impact on the fidelity and visibility of two-photon anti-coalescence. Thus the goal of providing universal deterministic pair separation becomes subject to the coupler design and performance. A coupler's response can depend largely on the input state, behaving as a 50:50 beamsplitter for some states, and as a wavelength demultiplexer for others. This in turn determines the extent to which quantum interference plays a role in tailoring the output state. Furthermore, separation performance can also depend on state properties such as the photon bandwidth and spectral entanglement. Many of these dispersion-related effects have no straightforward counterpart in the bulk optics paradigm.

\par Although dispersion may lead to a loss of interference, it can also serve to restore perfect splitting performance through WD effects. It is worth comparing this aspect to standard integrated WDs that function through purely classical means, as these may also be applied to a state in the form of Equation~(\ref{Eqn_StartingState_FullVers}) to provide splitting without collapsing spectral entanglement. Such devices typically use modal mismatch to restrict coupling to a narrow band of wavelengths near a single phase-matched point. This may suffice for a fixed non-degeneracy, yet high tunability in the photon wavelength separation serves an essential function for certain applications, such as emerging quantum photonic spectroscopy techniques \cite{Schlawin_NatureComms_2013, Schlawin_JChemPhys_2013}. In such cases the versatility of the interference-facilitated approach provides a clear advantage, especially if the tuning range must include degeneracy. For cross-polarized states, IFPS may also be contrasted against the merits of an integrated polarization-splitter (PS), where different coupling strengths for the $\mathrm{TM}$ and $\mathrm{TE}$ modes is critical. The PS can capitalize on waveguide dispersion and asymmetric geometries that are generally undesirable for 50:50 splitters, but of course cannot separate co-polarized states.

\section{Conclusions}
\label{Sec:Conclusions}

\par This paper examined coupler-mediated on-chip quantum interference as an avenue for deterministically separating photon pairs with arbitrary properties. General expressions were given for calculating bunched ($P_{\mathrm{B}}$) and anti-bunched ($P_{\mathrm{S}}$) outcome probabilities as well as associated interference visibilities ($V_{\mathrm{B}}$, $V_{\mathrm{S}}$). Several differences relative to the more familiar time-forwards HOM interference were also discussed. A theoretical investigation over the combined parameter space of the coupler and quantum state showed remarkable robustness of $P_{\mathrm{S}}$ against dispersion owing to natural compensation through wavelength-demultiplexing effects. The findings suggest that despite high levels of dispersion typical of integrated systems, interference-facilitated pair separation provides a promising monolithic solution for addressing the needs of highly tunable on-chip sources.

\section{Appendix}

\subsection{JSA Construction}
\label{Apdx:JSA_Construction}

\par For a given $\eta_{\sigma}\big(\xi_{\sigma}(\lambda; \lambda_{\mathrm{deg}})\big)$, the outcome probabilities $R_{pq}$ and interference visibilities $V_{pq}$ will depend on the properties of the two-photon state as characterized by the JSA. Rather than computing the JSA through device-specific mode dispersion parameters \cite{Chen_PRA_2005, Yang_PRA_2008}, it is more convenient here to define the JSA directly in terms of the photon bandwidths and central wavelengths of interest, to facilitate calculations that sweep over the photon parameters (e.g. $\Lambda$, $\Delta\lambda$). A JSA that mimics the output of a co-polarized (Type~I) SPDC process can be constructed in this way from the expression $\phi_{\sigma \sigma}(\omega_{1}, \omega_{2}) = \big[\zeta_{\sigma \sigma}(\omega_{1}, \omega_{2}) + \zeta_{\sigma \sigma}(\omega_{2}, \omega_{1}) \big]/\sqrt{2}$, where $\zeta_{\sigma \sigma}(\omega_{1}, \omega_{2}) = \phi^{(\text{P})}( \omega_{1} + \omega_{2})\phi^{(1)}_{\sigma}( \omega_{1})\phi^{(2)}_{\sigma}( \omega_{2})$ is a product of individual Gaussian spectra for the pump and twin photons. This construction satisfies the necessary exchange symmetry and has all the key qualitative features of a typical Type I JSA computed from SPDC theory.

\bibliographystyle{lpr}
\raggedright
\bibliography{journalTitlesAbbrv,RPM_MasterBib}

\end{document}